\documentclass[a4paper,fleqn,usenatbib]{mnras}
\pdfoutput=1
\usepackage[T1]{fontenc}
\usepackage{ae,aecompl}

\usepackage[usenames,dvipsnames]{color}
\usepackage{graphicx}	
\usepackage{amsmath}	
\usepackage{amssymb}	
\usepackage{orcidlink}
\usepackage[dvipsnames]{color}
\definecolor{tdgreen}{rgb}{0., 0.6, 0.}

\usepackage[english]{babel}
\usepackage[utf8]{inputenc}

\newcommand{\Msun}{\hbox{$\rm\thinspace M_{\odot}$}}

\usepackage{newtxtext,newtxmath}     

\title[Rapid X-ray Variability in Mkn~421]{Rapid X-ray Variability in Mkn~421 during a Multiwavelength Campaign}

\author[A.\ G.\ Markowitz et al.]
{Alex G.\ Markowitz,\orcidlink{0000-0002-2173-0673}$^{1,2}$\thanks{Email: \url{almarkowitz@camk.edu.pl}}
Krzysztof Nalewajko,\orcidlink{0000-0002-2019-9438}$^{1}$
Gopal Bhatta,\orcidlink{0000-0002-0705-6619}$^{3}$
Gulab C.\ Dewangan,\orcidlink{0000-0003-1589-2075}$^{4}$
\newauthor
Sunil Chandra,$^{5,6}$
Daniela Dorner,$^{7}$
Bernd Schleicher,$^{7}$
Urszula Pajdosz-\'Smierciak,$^{8}$
\newauthor
{\L}ukasz Stawarz,\orcidlink{0000-0001-8294-9479}$^{8}$
Staszek Zola,$^{8,9}$
Micha{\l} Ostrowski,$^{8}$
Daniele Carosati,$^{10,11}$
\newauthor
Saikruba Krishnan,\orcidlink{0000-0001-5375-9131}$^{1}$
Rumen Bachev,$^{12}$      
Erika Ben\'itez,{\orcidlink{0000-0003-1018-2613}}$^{13}$   
Kosmas Gazeas,\orcidlink{0000-0002-8855-3923}$^{14}$     
\newauthor
David Hiriart,$^{15}$     
Shao-Ming Hu,$^{16}$      
Valeri Larionov,$^{17,18}$ 
Alessandro Marchini,{\orcidlink{0000-0003-3779-6762}}$^{19}$
\newauthor
Katsura Matsumoto,{\orcidlink{0000-0002-5277-568X}}$^{20}$ 
A.A.\ Nikiforova,$^{17,18}$
Tapio Pursimo,$^{21}$      
Claudia M.\ Raiteri,\orcidlink{0000-0003-1784-2784}$^{22}$
\newauthor
Daniel E.\ Reichart,$^{23}$
Diego Rodriguez,$^{24}$   
Evgeni Semkov,$^{12}$     
Anton Strigachev,$^{12}$  
\newauthor
Yuki Sugiura,$^{20}$     
Massimo Villata,\orcidlink{0000-0003-1743-6946}$^{22}$  
James R.\ Webb,$^{25}$    
Axel Arbet-Engels,$^{26}$
\newauthor
Dominik Baack,$^{27}$ 
Matteo Balbo,$^{28}$ 
Adrian Biland,$^{26}$ 
Thomas Bretz,$^{26,29}$ 
\newauthor
Jens Buss,$^{27}$ 
Laura Eisenberger,$^7$ 
Dominik Elsaesser,$^{27}$ 
Dorothee Hildebrand,$^{26}$ 
\newauthor
Roman Iotov,$^7$ 
Adelina Kalenski,$^7$ 
Karl Mannheim,$^7$ 
Alison Mitchell,$^{26}$ 
\newauthor
Dominik Neise,$^{26}$ 
Maximilian Noethe,$^{27}$ 
Aleksander Paravac,$^7$ 
Wolfgang Rhode,$^{27}$ 
\newauthor
Vitalii Sliusar,$^{28}$ 
and Roland Walter$^{28}$ 
\\
$^{1}$Nicolaus Copernicus Astronomical Center, Polish Academy of Sciences, ul.\ Bartycka 18, 00-716 Warszawa, Poland\\
$^{2}$University of California, San Diego, Center for Astrophysics and Space
     Sciences, MC 0424, La Jolla, CA, 92093-0424, USA\\
$^{3}$Institute of Nuclear Physics, Polish Academy of Sciences, 31-342 Krak\'ow, Poland \\
$^{4}$Inter-University Centre for Astronomy and Astrophysics, Post Bag 4, Ganeshkhind, Pune 411007 India\\
$^{5}$Tata Institute of Fundamental Research, Mumbai, India\\
$^{6}$Centre for Space Research, North-West University, Private Bag X6001, Potchefstroom, 2520, South Africa\\
$^{7}$University of W\"urzburg, Institute for Theoretical Physics and Astrophysics, Emil-Fischer-Stra{\ss}e 31, Campus Hubland Nord, 97074, W\"urzburg, Germany\\   
$^{8}$Astronomical Observatory of the Jagiellonian University, ul.\ Orla 171, 30-244 Krak\'ow, Poland\\
$^{9}$Mt.\ Suhora Observatory, Pedagogical University, ul.\ Podchorazych 2, 30-084 Krak\'ow, Poland\\
$^{10}$EPT Observatories, Tijarafe, E-38780 La Palma, Spain \\
$^{11}$INAF, TNG Fundaci\'on Galileo Galilei, E-38712 La Palma, Spain \\
$^{12}$Institute of Astronomy and National Astronomical Observatory, Bulgarian Academy of Sciences, 72 Tsarigradsko Shosse Blvd., 1784 Sofia, Bulgaria\\
$^{13}$Universidad Nacional Aut\'onoma de M\'exico, Instituto de Astronom\'ia, AP 70-264, CDMX 04510, Mexico \\
$^{14}$Section of Astrophysics, Astronomy and Mechanics, Department of Physics, 
      National and Kapodistrian University of Athens, GR-15784, Zografos, \\ Athens, Greece \\
$^{15}$Instituto de Astronom\'ia, Universidad Nacional Aut\'onoma de M\'exico, AP106,  22860 Ensenada, BC, Mexico \\
$^{16}$Shandong Key Laboratory of Optical Astronomy and Solar-Terrestrial
     Environment, Institute of Space Sciences, School of Space Science and Physics,\\
     Shandong University, Weihai, Shandong, 264209, China \\
$^{17}$Astronomical Institute, St.\ Petersburg State University, 198504, St.\ Petersburg, Russia\\
$^{18}$Pulkovo Observatory, 196140, St.\ Petersburg, Russia \\
$^{19}$University of Siena, Department of Physical Sciences, Earth and Environment, Astronomical Observatory, Via Roma 56, I-53100 Siena, Italy\\
$^{20}$Osaka Kyoiku University, Institute of Astronomy, 4-698-1 Asahigaoka, Kashiwara 582-8582, Osaka, Japan \\
$^{21}$Nordic Optical Telescope, Apartado 474, E-38700 Santa Cruz de La Palma, Spain \\     
$^{22}$INAF, Osservatorio Astrofisico di Torino, via Osservatorio 20, I-10025 Pino Torinese, Italy \\
$^{23}$University of North Carolina at Chapel Hill, Dept.\ of Physics and Astronomy, Chapel Hill, NC, 27599, USA\\
$^{24}$Guadarrama Observatory, E-28430 Alpedrete, Madrid, Spain\\
$^{25}$Department of Physics, Florida International University, 11200 SW 8th St., Miami, FL, 33199, USA, and SARA Observatory\\
$^{26}$ ETH Zurich, Institute for Particle Physics and Astrophysics, Otto-Stern-Weg 5, 8093 Zurich, Switzerland \\
$^{27}$ TU Dortmund, Experimental Physics 5, Otto-Hahn-Str. 4a, 44227 Dortmund, Germany \\
$^{28}$ University of Geneva, Department of Astronomy, Chemin d'Ecogia 16, 1290 Versoix, Switzerland \\
$^{29}$ also at RWTH Aachen University \\    
}

\date{Accepted XXX. Received YYY; in original form ZZZ}

\pubyear{2021}

\begin{document}
\label{firstpage}
\pagerange{\pageref{firstpage}--\pageref{lastpage}}
\maketitle

\clearpage

\begin{abstract}  
The study of short-term variability properties in AGN jets has the potential to
shed light on their particle acceleration and emission mechanisms. 
We report results from a four-day coordinated multi-wavelength
campaign on the highly-peaked blazar (HBL) Mkn~421 in January
2019. We obtained X-ray data from AstroSAT, BVRI photometry with the
Whole Earth Blazar Telescope (WEBT), and TeV data from FACT to explore
short-term multi-wavelength variability in this HBL.
%
The X-ray continuum is rapidly variable on time-scales of tens of ks.
Fractional variability amplitude increases with energy across the
synchrotron hump, consistent with previous studies; we interpret
this observation in the context of a model with multiple cells whose
emission spectra contain cutoffs that follow a power-law distribution. 
We also performed time-averaged and time-resolved (time-scales of 6\,ks)
spectral fits; a broken power-law model fits all spectra well;
time-resolved spectral fitting reveals the usual hardening when
brightening behaviour.
Intra-X-ray cross correlations yield evidence for the 0.6--0.8~keV
band to likely lead the other bands by an average of $4.6\pm2.6$\,ks, but
only during the first half of the observation.
The source displayed minimal night-to-night variability at all wavebands 
thus precluding significant interband correlations during our campaign.
The broadband SED is modeled well with a standard one-zone leptonic model,
yielding jet parameters consistent with those obtained
from previous SEDs of this source.

\end{abstract}

\begin{keywords}  
BL Lacertae objects: Mkn~421 --
galaxies: active --
acceleration of particles --
black hole physics
\end{keywords}



\section{Introduction}

Blazars are a subset of Active Galactic Nuclei (AGN) whose luminosity is
dominated by emission from a relativistic jet lying close to the line of sight \citep{UP95}.
These jets emit non-thermal, Doppler-boosted emission across the
electromagnetic spectrum, with a characteristic broad two-hump
spectral energy distribution (SED). One hump, typically peaking in
IR-optical-UV, is synchrotron emission from relativistic electrons. The
other hump typically peaks in $\gamma$-rays, with leptonic (Inverse
Compton) and hadronic emission processes under consideration \citep{Mannheim93,Boettcher13}.
Classically, blazars have been broadly categorized 
based on optical spectral properties and on SEDs 
into the more luminous Flat-Spectrum Radio Quasars, with strong
emission lines, strong Compton dominance, and synchrotron emission typically 
peaking $\sim10^{13-14}$\,Hz, and BL Lac type objects (BL Lacs),
which are lower power and have
synchrotron emission peaking typically $\sim$ $10^{14-16}$\,Hz. 
BL Lacs can be further divided into high- and low-peaked BL Lacs (HBL/LBL) based on
synchrotron peak frequency.
More recent classification schemes for all blazars -- low-/intermediate-/high-peaked blazars --
are also based on SED peak frequencies \citep{Abdo10}. Strong continuum
variability on time-scales from minutes to years \citep[e.g.,][]{Ulrich97,Pian02,Bhatta16,BhDh20}
is observed across all blazar sub-classes.
Open issues to explore include determining the exact
location of the emission zones for each band 
and the dominant particle acceleration processes, e.g.,
propagating shocks, magnetic reconnection, and the role of magnetic
turbulence \citep{Mastichiadis97,Boettcher10,Yan13,Sironi15}.
Jets likely derive their power from rotational energy of spinning supermassive black holes \citep[e.g.,][]{BZ77,Sikora05},
though launching and collimation mechanisms remain unclear.
Insight into jet structure as well as emission mechanisms comes from
dedicated multi-wavelength (MWL) continuum flux and polarization
monitoring campaigns that enable us to search for interband
correlations and model the time-dependent behaviour of the SED,
particularly towards relatively short time-scales.

 %
In HBLs, X-ray and TeV are thought to both probe the highest-energy
population of electrons, as supported by commonly-observed
correlations between X-ray and TeV bands, as well as variability
amplitude increasing with energy within each SED hump.
Mkn~421 is one of the most famous HBL/HSP blazars: it was the first
extragalactic-confirmed TeV source \citep{Punch92}, and has been
targeted by MWL campaigns of varying spectra/temporal coverage and
involving a wide range of space- and ground-based telescopes.  Its
flux density can span up to two orders of magnitude in many bands, though
it is known to usually be bright at optical, X-ray, and TeV wavebands,
and frequently displays rapid ($<$1--2\,day) variability in the X-ray
and TeV wavebands \citep[e.g.,][]{Blazejowski05,Giebels07,Abdo11,Aleksic15b,Balokovic16}.

Variability as short as minutes--sub-hours in
X-ray and $\gamma$-rays has been observed in Mkn~421 and other HBLs, suggesting these bands
probe the most energetic electrons, which have
shortest cooling times \citep[e.g.,][]{Aharonian07,Ackermann16}.  In Mkn~421, X-ray/TeV
emission is generally correlated on time-scales of week/months/years,
usually at zero time lag, across multiple flux levels
\citep[e.g.,][]{Amenomori03,Rebillot06,Albert07,Aleksic15b,Balokovic16,Ahnen16,Arbet21},
though \citet{Fossati08} observed a $2.1 \pm 0.7$\,ks X-ray-to-TeV lag
during a strong flare in 2001.  Such strong correlations are also
observed in other HBLs
\citep[e.g., Mkn\,501,][]{Krawczynski00,Krawczynski02}.  Combined with SED modeling,
these correlations generally support a one-zone synchrotron
self-Compton (SSC) scenario \citep[e.g.,][]{Acciari11,Aleksic12,Abeysekara17}, though
hadronic models may also be relevant \citep{Boettcher13}.

Meanwhile, optical to TeV/X-ray correlations are only sometimes seen
in HBLs, and are not a common property; the optical band
probes lower electron energies ($\nu_{\rm syn} \propto \gamma^2$),
and the optical emitting region likely has a higher volume filling factor than X-ray/TeV.
Furthermore, some campaign results on Mkn~421 support multiple compact
regions, e.g., two-zone SSC models: Correlated X-ray/TeV flares,
poorly correlated with optical emission, observed during a 13-day MWL
campaign in 2010 were fit by two distinct blobs (\citealt{Aleksic15b};
see also \citealt{Balokovic16}).  Similarly, observations of optical
polarization degree (PD) not correlating simply with flux or wide
rotations of position angle (PA) also support multiple emission
regions \citep{Carnerero17} with different volumes and/or differing
electron energy distribution (EED) widths.
An additional complication is the occasional observation of TeV orphan
flares in Mkn~421 \citep[and other HBLs, e.g.,][]{Acciari11},
or an X-ray orphan flare \citep[e.g.,][]{Rebillot06,Lichti08,Acciari20}; such flares
mandate either fine-tuning of SSC models \citep{Krawczynski04},
alternate models such as a spine/sheath configuration \citep{Ghisellini05},
or gamma rays produced by hadrons \citep[e.g., proton-synchrotron emission,][]{Muecke01}.


We thus organized a coordinated multi-wavelength campaign in X-rays,
TeV, and optical/NIR occurring 2019 January 10--14, MJD 58493 -- 58497.
This was a campaign
planned in advance, and not organized in response to any target of opportunity alert.
Our strategy here is to concentrate on short time-scale variability
characteristics in Mkn~421, using a continuous 4-day campaign.
Our science goals are to build up a more complete physical picture of
the jet, including the properties of emitting regions.
%
%
%
%

The rest of the paper is organized as follows: $\S2$ describes the
observations and data reduction, including extraction of light curves
and spectra.  In $\S3$, we explore continuum variability within
separate bands, including variability amplitudes.  Cross-correlations
are presented in $\S4$.  In $\S5$, we present the results of
time-averaged and time-resolved X-ray spectral fits.  The FACT TeV
spectra are presented in $\S6$, and we present our broadband SED in
$\S7$.  We present a discussion of the results in $\S8$, and a summary
of our main conclusions in $\S9$.

\section{Observations and Data Reduction}

\subsection{AstroSat}

\textit{AstroSat} \citep{Singh14} is the first dedicated Indian astronomy mission aimed
at studying celestial sources in X-ray, UV and limited optical
spectral bands simultaneously.
It was launched from Satish Dhawan Space Centre, Sriharikota on 28
September 2015 into a 650 km near-equatorial orbit with 6-degree
orbital inclination; this orbital inclination helps to minimize
(though not eliminate) passage through the South Atlantic Anomaly on
most orbits.  \textit{AstroSat} is operated by the Indian Space Research
Organisation (ISRO).

\textit{AstroSat} contains five payloads spanning the optical/UV and soft--hard
X-ray ranges; in this paper, we use data from the Soft X-ray Telescope
(SXT) and Large Area Xenon Proportional Counters (LAXPC).
Unfortunately, we could not use the Ultraviolet Imaging Telescopes to
observe Mkn~421 in the UV range; an extremely bright star in the UVIT
field of view would have violated the instrument's safety threshold.

\textit{AstroSat} continuously observed Mkn 421 from 07:47 UT on 2019 January 10
to 18:59 UT on 2019 January 14, for a duration of 68 orbits (orbit
numbers 17764--17831), with time lost only to Earth occultation
occurring every 97 minutes.

\subsubsection{SXT reduction}

The SXT\footnote{\url{https://www.tifr.res.in/~astrosat_sxt/index.html}} \citep{Singh17}
is a focusing telescope with conical foil mirrors and an X-ray CCD
detector; it covers the 0.3--7\,keV range.  Level 1 data were
reprocessed to produce cleaned level 2 events; to do this, we used SXT
software version
1.4b\footnote{\url{www.tifr.res.in/~astrosat_sxt/sxt_pipeline.php}},
including the GTI corrector and
event merger Julia script \textsc{SXTEVTMERGERTOOL}\footnote{\url{http://www.tifr.res.in/~astrosat_sxt/dataanalysis.html}}.
We used Xselect v.~2.4g in HEASOFT v.\ 6.26 to extract a
time-averaged spectrum; we extracted counts from a 16$\arcmin$ radius
circle centered on the source. We also extracted time-resolved
spectra, one for each orbit.  We extracted background-subtracted light
curves over the 0.6--7\,keV band and the 0.6--0.8, 0.8--1.1, 1.1--1.4,
1.4--1.9, 1.9--3.0, and 3.0--7.0\,keV sub-bands, which we then binned to the
satellite orbit (5830\,s); we did not detect significant variability
on shorter time-scales. 
A blank-sky SXT spectrum (SkyBkg\_comb\_EL3p5\_Cl\_Rd16p0\_v01.pha), provided by the SXT
team, was used to estimate background count rates within each sub-band.
In all spectral fits, we use the ``gain''
command to modify the response and alleviate calibration
issues associated with modeling the instrumental Si K and Au M edges
near $\sim1.8$ and $\sim2.2$\,keV, respectively; the gain slope was fixed to 1.
All SXT spectra were fit over the 0.6--7.0\,keV range.
We used the latest response and effective area files provided by the SXT team
for sources observed on-axis: sxt\_pc\_mat\_g0to12.rmf and sxt\_pc\_excl00\_v04\_20190608.arf.

\subsubsection{LAXPC reduction}

The LAXPC consists of three large-area ($\sim6000$~cm$^{2}$) X-ray proportional counter units,
each with a time resolution of 10 $\mu$s
and covering the 3--80\,keV range \citep{Yadav16,Agrawal17}.
For this observation, we used data from LAXPC unit 2 only.
Unit 3 was switched off in March 2018 due to abnormal gain changes. Unit 1
underwent high voltage adjustments in Spring 2018, and so standard response files
cannot be applied; we may include unit 1 data in a future
analysis when updated response files are made available by the LAXPC Team.

We used the 2018~May~19 Format~A
version of the LAXPC software\footnote{\url{http://astrosat-ssc.iucaa.in/?q=laxpcData}},
and we reprocessed level 1 events using \textsc{laxpc\_make\_event}
to generate cleaned level 2 event files.
We refer the reader to \citet{Antia17} for further details on response matrix generation
and model-based estimation of background spectra.
We extracted a time-average spectrum, as well as time-resolved spectra, 
one for each orbit; again, no significant variability
on shorter time-scales was confirmed. 
Spectra were initially extracted in 1024 channels but rebinned to an energy
resolution of 5\,per cent (three times higher than
the LAXPC energy resolution).
We also extracted background-subtracted and orbitally-binned
light curves over the 3--7, 7--12, and 12--20\,keV bands.
The 3--7\,keV LAXPC light curve agreed well with the 3--7\,keV SXT light curve in terms of their variability trends.
However, while those trends were also present in the 7--12 and 12--20\,keV bands,
these latter two bands showed additional variability trends not observed $<7$\,keV.
These additional trends matched very well with 
variability trends detected at $>40$\,keV, where the source is not significantly detected,
and are thus likely artefacts of background modeling and subtraction.
For example, the 12--20\,keV net and background count rates are 3.2 and 12.3\,ct s$^{-1}$, respectively;
$\sim2$\,per cent background fluctuations (the expected level of systematic uncertainty in
background modeling, e.g., \citealt{Antia17}) would cause spurious fluctuations of up to $\sim8$\,per cent
in the net light curve, consistent with the observed net light curve.
We thus avoided using the $>7$\,keV light curve data for this object,
and restricted spectral fitting to $<7$\,keV.
In all spectral fits, we use the ``gain'' command to modify
the response and alleviate calibration issues, with the gain slope
fixed to 1.

\subsection{FACT}

The First G-APD Cherenkov
Telescope\footnote{\url{https://fact-project.org/}} (FACT),
an imaging air-Cherenkov
telescope located on the Canary Island La Palma in Spain, is monitoring
blazars at TeV energies \citep{2013JInst...8P6008A}. Using silicon-based
photosensors \citep{2014JInst...9P0012B}, the duty cycle of the
instrument is maximized while at the same time the gaps in the light
curves are minimized. Mkn\,421, Mkn\,501, 1ES\,1959+650 and
1ES\,2344+51.4, the brightest blazars at TeV energies, are observed on
a nightly basis within the visibility windows for about 120 to 200
nights per year. A total of up to 2600~hours of physics data are
recorded per year \citep{2019ICRC...36..665D}. 

Usually the blazars are observed between 40~minutes and 6~hours per
night depending on their visibility, where the schedule is maximized
for best threshold. However, for the January 2019 campaign on Mkn 421,
the observation windows were maximized for the 5 nights of the
campaign (2019 January 10 -- 14, morning UTC time) to increase the
simultaneous coverage with {\it AstroSat} and WEBT.

The data were analysed using the Modular Analysis and Reconstruction
Software (MARS) \citep{2010apsp.conf..681B} with the low level analysis
as described in \citet{2015arXiv150202582D}. For the background
suppression, the ``light curve cuts'' and the ``spectrum cuts'' as
described in \citet{2019ICRC...36..630B} were applied correspondingly
to calculate the light curves and spectra. To correct the light curves
for observational effects, the dependence of the gamma ray rate from
zenith distance and trigger threshold (changing with ambient light
conditions) was determined using the gamma ray rate measured from the
Crab Nebula, a standard candle at TeV energies. Correction factors were
applied to the light curves as described in \citet{Arbet21}.

To select good quality data, a selection cut based on the cosmic-ray
rate \citep{2019ICRC...36..630B,2017ICRC...35..779H} using the
artificial trigger rate $R750$ was applied. Choosing a threshold of 750
DAC-counts, the dependence of $R750$ on the zenith distance was studied
and a corrected rate $R750_{\rm cor}$ was determined as described in
\citet{2017ICRC...35..612M} and \citet{2019APh...111...72B}. To take
into account seasonal changes of the cosmic-ray rate due to changes in
the Earth's atmosphere, a reference value $R750_{\rm ref}$ was determined
per moon period. Good quality data were selected using $0.93 <
R750_{\rm cor}/R750_{\rm ref} < 1.3$. 
After data quality selection, the following data sampling is obtained:
4.9, 5.5, 5.0, 4.4, and 4.5\,hours for
10, 11, 12, 13, and 14 Jan., respectively.

\subsection{WEBT}

The Whole Earth Blazar
Telescope\footnote{\url{https://www.oato.inaf.it/blazars/webt/}}
(WEBT) is a global network of observatories which routinely coordinate to
conduct uninterrupted campaigns on blazars to explore their short
time-scale variability characteristics \citep[e.g.,][]{Villata2002, Raiteri2017}.
Sixteen observatories in the WEBT 
consortium contributed to this campaign (listed in Table~\ref{tab:participating}).  The target was observed in
four wide band Johnson-Cousins photometric system filters: $B$, $V$, $R$, and $I$ (where available), with
effective central wavelengths of 4353, 5477, 6349, and 8797\,\AA, respectively.
Exposure times for individual images ranged from
30 to 1600 s, depending on the telescope parameters.

The standard
calibration method has been applied to all of the acquired images
(bias, dark and flat-field images correction). The Skynet telescopes'
raw data were calibrated with the software provided by The Skynet
Robotic Telescope Network \citep{Zola21},
while for the rest of the images we used
Image Reduction \& Analysis Facility
(IRAF)\footnote{\url{http://ast.noao.edu/data/software}}. The C-MUNIPACK
package\footnote{\url{http://c-munipack.sourceforge.net/}} was used to
perform differential aperture photometry by a single person (U.P.-S.)
to ensure uniformity in the results. The comparison star was at
RA = $11^{\rm h}~04^{\rm m}~51\fs14$, Dec=$+38^{\circ}~17\arcmin~10\farcs8$ from \citet{Villata98};
the check stars are
located at
RA = $11^{\rm h}~04^{\rm m}~18\fs18$, Dec=$+38^{\circ}~16\arcmin~31\farcs1$ from \citet{Villata98} or
RA = $11^{\rm h}~04^{\rm m}~08\fs48$, Dec=$+38^{\circ}~22\arcmin~26\farcs5$ from \citet{McGimsey76} (all are epoch J2000).
To achieve consistent results, we chose the radius of
photometry aperture based on the pixel scale of each instrument's CCD
camera. Because of bad quality, some data have been excluded from
further analysis.

For the purpose of analysis of
fractional variability amplitudes, we
selected the observations from the EPT Observatory, La Palma;
this observatory's $V$, $R$, and $I$ band
light curves had the best sampling -- continuously for durations
ranging from 15.5 to 23.4\,ks (4.3 to 6.5\,hours) each night for four consecutive nights
-- and yielded the best variability-to-noise ratio.
For the purpose of constructing the broadband SED, however, we used
all data to compute time-averaged magnitudes in $B$, $V$, $R$, and $I$ bands,
and converted to flux densities using zeropoints from \citet{Bessell1998}.

We corrected the derived flux densities for optical extinction due to
the Galaxy, $N_{\rm H, total} = 2.03 \times 10^{20}$ cm$^{-2}$
\citep{Willingale13xx}, and using the dust/gas extinction ratio of
\citet{Nowak12}, $N_{\rm H} = A_{\rm V} \times 2.69\times10^{21}$\,cm$^{-2}$~mag$^{-1}$, which yields $A_{\rm V}= 0.05$\,mag.
Following the Galactic extinction curve of \citet{Cardelli89},   
$A_{\rm B} \sim 1.2 A_{\rm V}$ = 0.06\,mag,
$A_{\rm R} \sim 0.8 A_{\rm V}$ = 0.04\,mag, and
$A_{\rm I} \sim 0.6 A_{\rm V}$ = 0.03\,mag.
The $B$, $V$, $R$, and $I$ band flux densities were multiplied by 1.06, 1.05, 1.04,
and 1.03, respectively.

We estimated and subtracted the contamination from the elliptical host galaxy:
\citet{Carnerero17} estimated the R-band contamination in the host galaxy to be 7.86\,mJy for a 5~arcsec aperture.
Following a de Vaucouleurs profile (Eqn.~1 in \citealt{Carnerero17}), we estimate that a 4~arcsec aperture
would see a contamination level of 5.37\,mJy, which is 33\,per cent of the total (jet + host galaxy) flux observed during our campaign.
We applied the colour indices of \citet{Mannucci01} for elliptical
galaxies ($B-V= 0.99\pm0.05$, $V-R=0.59\pm0.05$, and
$V-I=1.22\pm0.07$) to estimate the level of contamination at $B$ band
(1.66\,mJy; 17\,per cent of total observed flux), $V$ band (3.70\,mJy; 26\,per cent of
total), and $I$ band (7.57\,mJy; 38\,per cent of total).  Fractional
variability amplitudes (defined and calculated in $\S$\ref{sec:calcfvars}) for $V$, $R$,
and $I$ bands were subsequently increased by factors of 1.35, 1.50, and
1.61, respectively.

The final time-averaged flux densities for the jet at $B$, $V$, $R$, and $I$
bands were 7.99, 10.45, 10.70, and 12.41 mJy, respectively; fluxes in
units of erg~cm$^{-2}$~s$^{-1}$ are listed in Table~\ref{tab:fvar}.

\begin{table*}
        \centering
        \begin{tabular}{lccccc} \hline
          \multicolumn{6}{c}{\textbf{Contributing Optical Observatories}}  \\ 
                       & \multicolumn{4}{c}{Number of Exposures} &  Dates   \\ \hline
Observatory/Telescope  &  $B$ & $V$ & $R$ & $I$  &                (JD $-$ 2458000.0)  \\ \hline
EPT Observatory, 40 cm, Tijarafe, La Palma, Spain     &     & 76 & 89  & 88   &   494, 495, 496, 497\\

Univ.\ of Athens Obs.\ (UOAO), 40 cm, Greece          &     &    & 314 &      &  492, 499, 501, 502 \\  

Astron.\ Obs., Univ.\ of Siena, Italy                 &  36 & 47 & 34  & 73   &   495, 498 \\  

Osaka Kyoiku Univ.\ Obs., 51 cm, Japan                &     & 35 & 38  & 38   &   497, 498\\ 

NOT-ALFOSC, La Palma                                  &     & 56 &     &      &   496  \\    

SARA/Kitt Peak, 90 cm                                 &     &   & 40  &  & 494   \\ 

PROMPT-5, Cerro Tololo Inter-American Univ., Chile    & 11  & 13 &  9  & 14   &  493, 494, 495, 496, 498\\

PROMPT-8, Cerro Tololo Inter-American Univ., Chile    &  3  & 12 &     & 15   &  494, 495, 497   \\

Weihai Obs.\ of Shadong Univ., 100 cm, China          &  4  &  8 &  9  &      &  498  \\

Guadarrama Observatory, 25 cm, Alpedrete, Madrid, Spain &  &  7  & 6  &  4  &  496, 497 \\ 

PROMPT-MO-1, Meckering Obs., Australia                &  2  &  5 &  2  &  8   &  496, 497, 498 \\

RRRT, Fan Mountain Observatory, Virginia, USA         &  2  &    &  6  &  8   &  494 \\

St.\ Petersburg Univ., 40 cm, Russia                  & 3 & 3 & 3 & 4 & 496  \\ %

Astron.\ Obs.\ Belgrade, 60 cm, Serbia                &     &  4 &  4  &  4   &   488, 491   \\   

Inst.\ of Astron.\ and NAO, Rozhen Obs., 50/70 cm, Bulgaria &  2  &  2 &  2  &  2   &  496 \\

San Pedro Mart\'ir Observatory, 84 cm, Mexico         &     &    &   2 &      &  495, 496 \\   


\hline
\end{tabular}
        \caption{The WEBT observatories/telescopes contributing to the January 2019 campaign.   }
\label{tab:participating}
\end{table*}

\subsection{\it Fermi}

We extracted a contemporaneous GeV spectrum using data from the
\textit{Fermi} Large Area Telescope (LAT; \citealt{Atwood09}).
Data were analysed with the {\tt ScienceTools}
software package ({\tt v11r5p3}) using the instrument response
function (IRF) {\tt P8R2\_SOURCE\_V6}. LAT events were selected from the
region of interest (ROI) of radius $10^\circ$ centered at Mkn~421. We
have established that for events detected over the time period of
${\rm MJD} = 58493-7$ with reconstructed arrival direction $<2^\circ$
from Mkn\,421, the maximum reconstructed energy was $7.9\;{\rm GeV}$,
hence we restricted the analysis to the energy range of $(0.1 -
8)\;{\rm GeV}$. We applied selection cuts for good time intervals
(GTI), avoiding the South Atlantic Anomaly (SAA), using 
zenith angles
$<100^\circ$. Our source model includes sources from the 3FGL catalog
\citep{Acero15} located within $20^\circ$ from
Mkn~421, but only those that had been detected over a longer time
period of ${\rm MJD} = 58450-58550$ with test statistic ${\rm TS} > 1$;
it also includes
the Galactic background template {\tt gll\_iem\_v06}, and the
isotropic background spectrum {\tt iso\_P8R2\_SOURCE\_V6\_v06}. In the
maximum likelihood analysis, the spectrum of Mkn\,421 was fitted with
the {\tt PowerLaw2} model.

We did not extract a light curve (e.g., daily-binned), as the
variability-to-noise ratio was too low for meaningful constraints on
variability amplitudes or interband correlations.

\section{Overview of Multiband Variability}

\begin{figure*}
  \includegraphics[width=1.8\columnwidth]{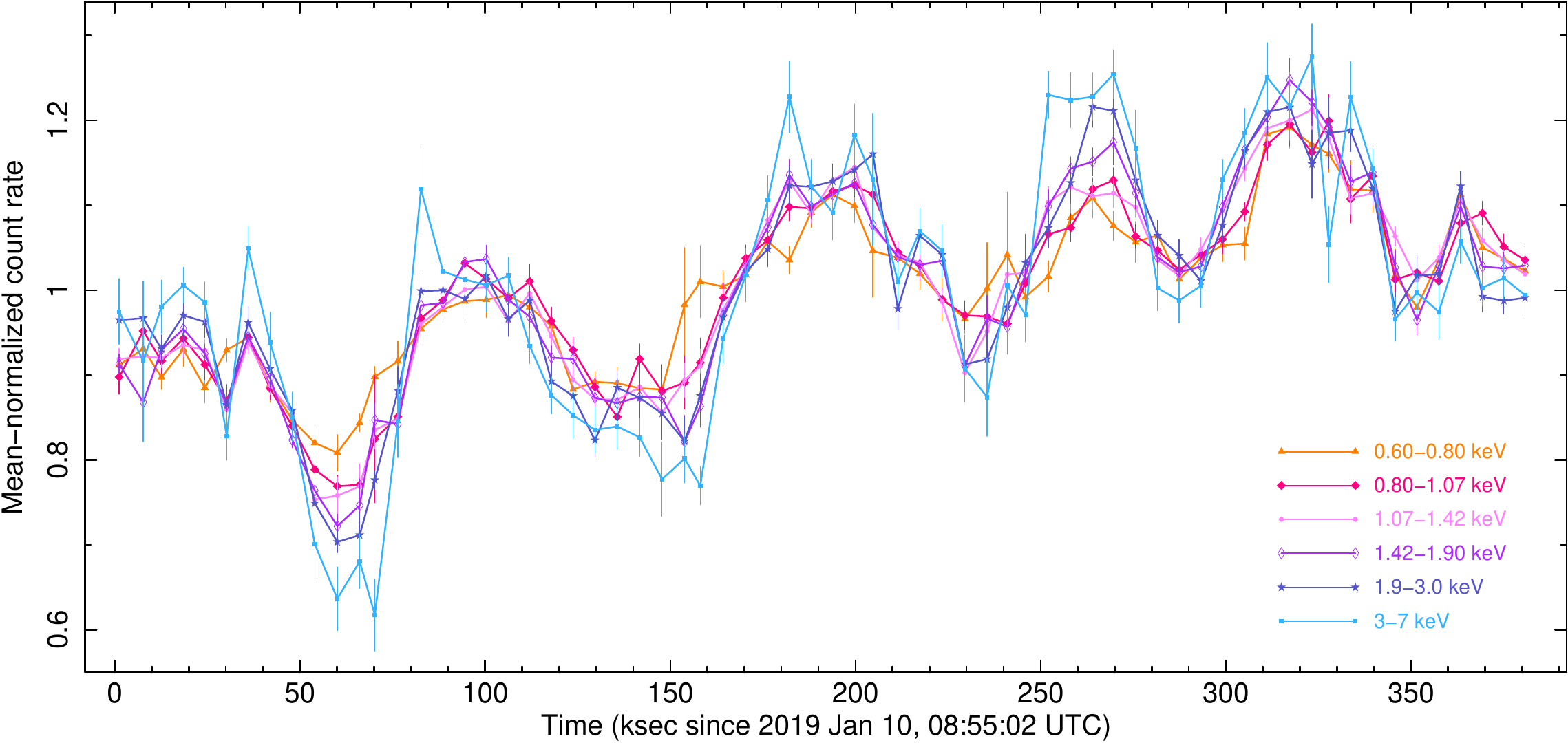}
    \caption{Mean-normalized, orbitally-binned \textit{AstroSAT} SXT light curves.}
    \label{fig:SXTlcoverplot}
    \end{figure*}

SXT sub-band light curves, normalized by their means, are presented in
Fig.~\ref{fig:SXTlcoverplot}.
There is a quasi-sinusoid-like trend dominating the variability in our
campaign.  However, given the small number of cycles observed and the
known propensity of red noise variability processes in all categories
of AGN, such a trend does not constitute evidence of a quasi-periodic
signal and is consistent with a pure red noise process
\citep[e.g.,][]{Vaughan16}, as discussed
further in Appendix A.

It is additionally curious that the first couple periods of this
quasi-sinusoid-like trend are in the neighborhood of 90\,ks ($\sim$1\,day).  We thus
explored all possible satellite and detector housekeeping parameters
to exclude any possible instrumental effect.  We found no correlation
with satellite attitude, satellite passage through the South Atlantic
Anomaly, charged particle background rate (as measured by the on-board
Charged Particle Monitor), satellite pointing offsets or roll angle,
quality grade of SXT operating conditions,  nor any detector
operational parameters (e.g., CCD and camera temperatures).  We
conclude that this trend is intrinsic to Mkn~421.

Following e.g., \citet{Balokovic16}, we characterized the typical
variability time-scale by fitting each rise/fall in the 1.1--1.4\,keV
light curve (which has the highest count rate of the SXT sub-bands)
with an exponential rise/decay, e.g., $F(t) = e^{-t/\tau_{\rm var}}$.
We find the mean $\tau_{\rm var}$ in this band to be 130\,ks.



\begin{figure*}
  \includegraphics[width=1.8\columnwidth]{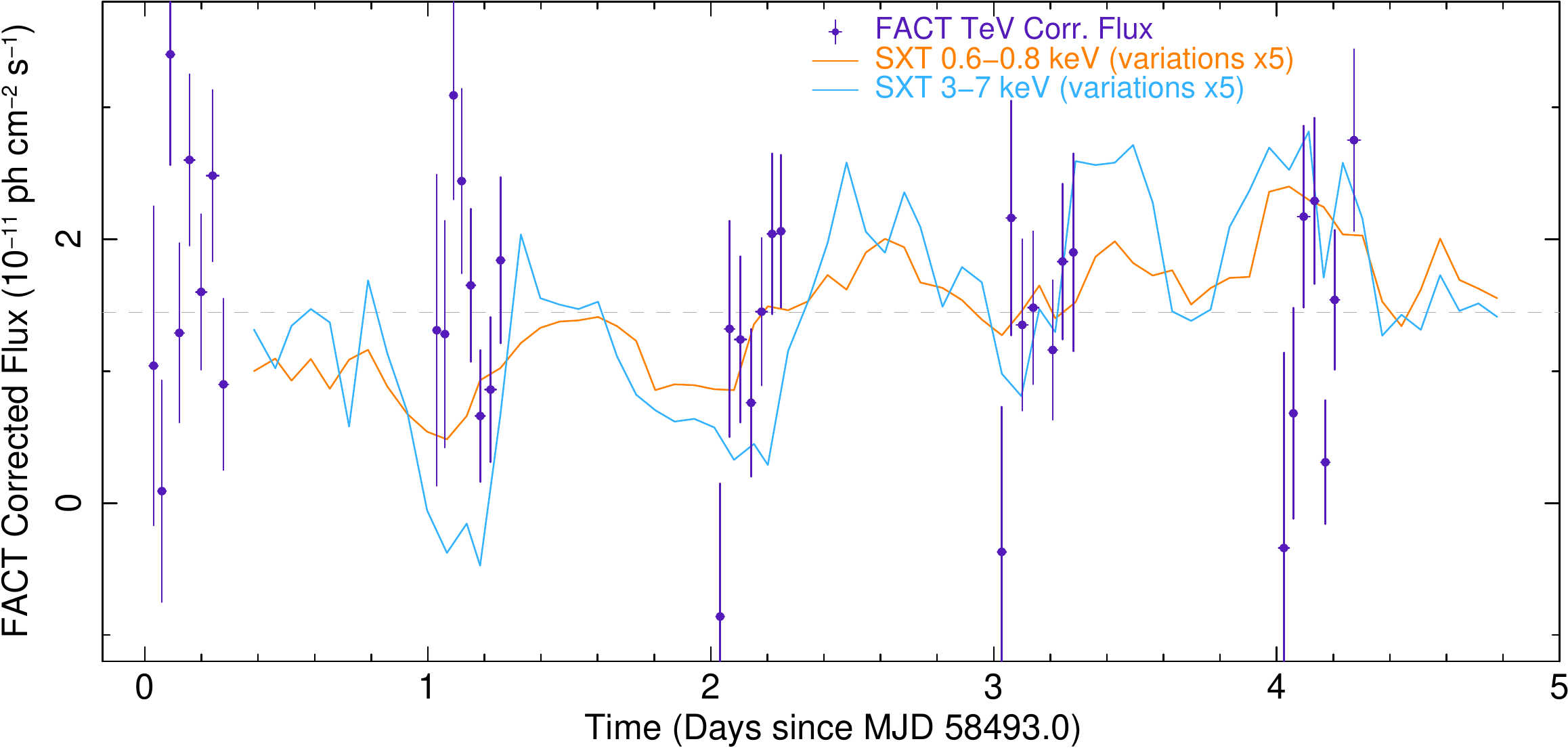}
  \caption{Mean-normalized FACT $>$0.7 TeV corrected flux light curve, 40-minute bins (purple data points).
    Overplotted are the \textit{AstroSat} SXT 0.6--0.8\,keV (orange line) and 3--7\,keV (cyan line) light curves,
      re-normalized to the mean of the TeV light curve ($1.44 \times10^{-11}$ ph cm$^{-2}$ s$^{-1}$); variability
      relative to the mean in the SXT light curves has been amplified
      by 5 for visualization purposes only.  }
    \label{fig:FACTSXToverplot}   

    \end{figure*}

The $>0.7$~TeV FACT corrected flux light curve is presented,
overplotted with re-normalized SXT light curves, in
Fig.~\ref{fig:FACTSXToverplot}.  Neither the TeV nor the X-ray band
shows strong day-to-day variation during the duration of our campaign.
though there are TeV variability trends observed on $\sim$hour
time-scales on some nights.
Potential X-ray/TeV interband correlations are explored further in
$\S$\ref{sec:xtzerolag}.

\begin{figure*}
  \includegraphics[width=1.8\columnwidth]{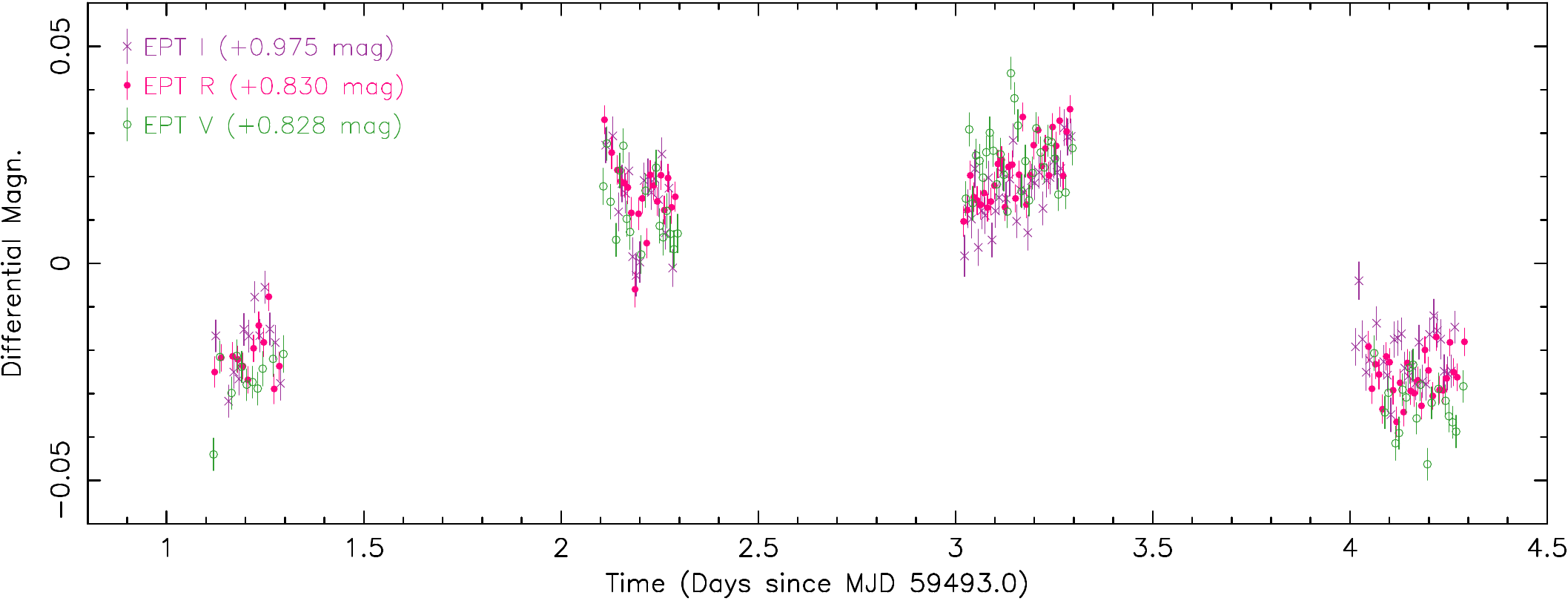}
  \caption{I-, V-, and R-band photometry from the EPT telescope, renormalized to a common differential magnitude.}
    \label{fig:EPTIRVoverplot}
    \end{figure*}

The I-, R-, and V-band differential magnitude photometric data points
obtained at the EPT telescope are plotted in
Fig.~\ref{fig:EPTIRVoverplot}.  The maximum/minimum flux ratios are
quite small, only $\sim0.3$\,per cent in I-band and $\sim0.6$\,per cent in R-band, and not atypical for
Mkn~421.





\subsection{Variability amplitudes}\label{sec:calcfvars}

\begin{figure}
  \includegraphics[width=0.8\columnwidth]{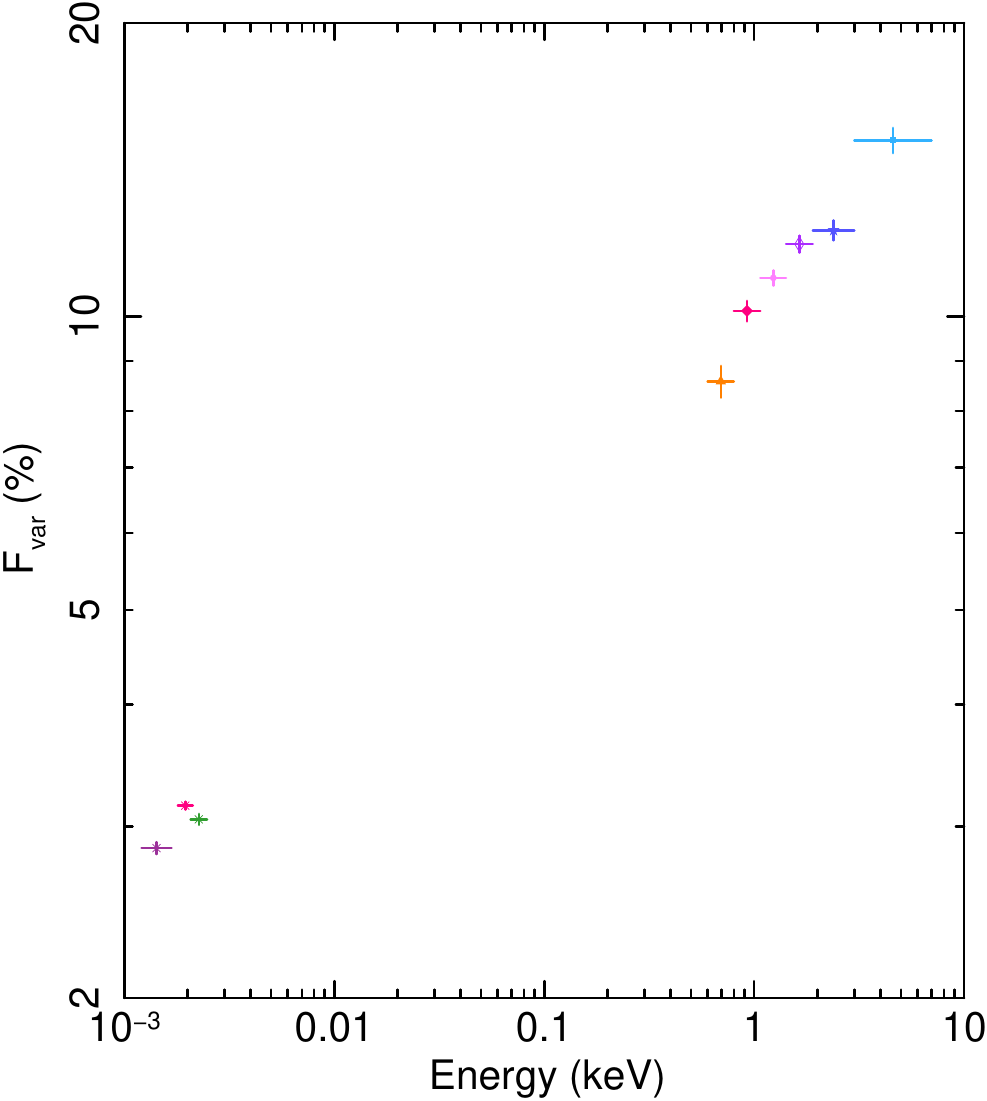}
    \caption{Fractional variability amplitude $F_{\rm var}$ increases monotonically
      as a function of photon energy across the \textit{AstroSat} SXT bands, and increases with photon energy
      from the NIR/optical bands to the X-rays.}
    \label{fig:FVAROX}
    \end{figure}

We calculated fractional variability amplitude $F_{\rm var}$ \citep{Vaughan03} for each \textit{AstroSat} SXT band
and for the WEBT/EPT I-, R-, and V-bands (in flux units, not magnitudes); the results are listed in Table~\ref{tab:fvar}.
At $\sim9-15$\,per cent, our values are bit lower than
typical values for $F_{\rm Var}$ in
X-rays on timescales of $\sim$days \citep[$\sim40-80$\,per cent, e.g.,][albeit at different flux states]{Giebels07,Balokovic16,Acciari20}.
Within the X-ray band,
$F_{\rm var}$ increases monotonically with photon
energy; $F_{\rm var}$ also increases with energy from the NIR/optical ot the X-rays, as illustrated in Fig.~\ref{fig:FVAROX}.
We find best-fitting relations (in log-log space) of $F_{\rm var}$ $\propto$ $E^{0.20\pm0.01}$ (NIR/optical + X-ray)
or $F_{\rm var}$ $\propto$ $E^{0.26\pm0.02}$ (X-ray only).

Our results are qualitatively consistent with previous measurements of
$F_{\rm var}$ increasing with energy in HBL blazars across the X-ray
band (\citealp[e.g., references in][]{Ulrich97,Kataoka00}), and
qualitatively consistent with measurements of $F_{\rm var}$ increasing
within each SED hump in Mkn~421 and Mkn~501 \citep{Giebels07,
  Balokovic16, Schleicher19,Acciari20,Arbet21}, 
although \citet{Acciari20} noted such behaviour occurring in
most, but not all, of their observations.
Our results are qualitatively consistent with \citet{Giebels07}, who
measured $F_{\rm var}$ $\propto$ $E^{1/4}$ from optical to UV to
X-rays and with \citet{Fossati00}, who found $F_{\rm var}$ $\propto$
$E^{\sim 1/4}$ across the X-ray band. We must note
  caveats against such direct comparisons of $F_{\rm var}$, however,
  given the slightly different window functions: Here, our light
  curves have durations of $\sim$4\,days, while the durations of
  \citet{Giebels07} and \citet{Fossati00} are 1--7\,days and 2.3\,days,
  respectively.  In addition, as noted by \citet[][their
    $\S$4.3.1]{Fossati00}, there may very well be different flux
  probability distributions between different campaigns; $F_{\rm var}$
  also implicitly assumes a Gaussian distribution, which does not
  apply to typical blazar light curves \citep[e.g.,][]{Emman13}.

  In the TeV band, using 40 minute-binned light curves,
  and not omitting the negative-flux points,
we measure $F_{\rm var} =  38.9 \pm 12.1$\,per cent.
This value is somewhat low compared to what others have measured
over timescales of $\sim$days in the TeV range
\citep[e.g., very roughly 40 per cent to above 100 per cent][]{Balokovic16,Acciari20}.

\begin{table*}
        \centering
        \begin{tabular}{llcc} \hline
          \multicolumn{4}{c}{\textbf{Variability Amplitudes and Mean Count Rates or Fluxes}}  \\ \hline
Bandpass & Instrument &  Mean   & $F_{\rm var}$ (per cent) \\ \hline
$I$ & WEBT & $4.67\times10^{-11}$ erg cm$^{-2}$ s$^{-1}$   & $2.85  \pm 0.04$ (EPT only) \\
$R$ & WEBT & $5.03\times10^{-11}$ erg cm$^{-2}$ s$^{-1}$   & $3.15  \pm 0.03$ (EPT only) \\
$V$ & WEBT & $5.75\times10^{-11}$ erg cm$^{-2}$ s$^{-1}$   & $3.05  \pm 0.04$ (EPT only) \\
$B$ & WEBT & $5.43\times10^{-11}$ erg cm$^{-2}$ s$^{-1}$   &           \\
0.6--0.8 keV & SXT & 1.47 ct s$^{-1}$ & $ 8.58 \pm 0.32$ \\ 
0.8--1.1 keV & SXT & 2.36 ct s$^{-1}$ & $10.33 \pm 0.25$ \\ 
1.1--1.4 keV & SXT & 2.71 ct s$^{-1}$ & $10.96 \pm 0.20$ \\ 
1.4--1.9 keV & SXT & 2.22 ct s$^{-1}$ & $11.87 \pm 0.24$ \\
1.9--3.0 keV & SXT & 1.31 ct s$^{-1}$ & $12.26 \pm 0.29$ \\
3--7    keV  & SXT & 0.67 ct s$^{-1}$ & $15.16 \pm 0.45$ \\ 
0.7--11 TeV   & FACT (40-min.\ bins) & $1.44 \times10^{-11}$ ph cm$^{-2}$ s$^{-1}$  &  $38.9 \pm 12.1$  \\  \hline
\end{tabular}
        \caption{Variability amplitudes measured over the full duration of each light curve.
          The third column refers to flux in the case of WEBT (all observatories with good data),
          mean count rate in the case of \textit{AstroSat} SXT, and mean TeV corrected flux in the case of FACT.
          Optical fluxes and
          measurements of $F_{\rm var}$ have been corrected for Galactic foreground absorption and host galaxy contamination.
          $F_{\rm var}$ was calculated in flux units for the WEBT (EPT-only) light curves.
          }
\label{tab:fvar}
\end{table*}


\subsection{Fitting of individual X-ray flares}\label{sec:fitflare}


We divided each sub-band light curve into four separate flares, each one encompassing a rise and a decay.
We ignored the first 67\,ks of data (decay only), and define four
flare start/stop times within each sub-band
light curve by the minima closest to 67, 150, 240, 295, and 355\,ks since the start of the observation.

For each separate flare, we removed the baseline ``slope'' by
subtracting a linear trend connecting the end points;
we did not fit with an additional linear component with free parameters
as that would have added two additional free parameters.
We then fit a model following \citet[][their Eq.\ 5]{Acciari20}:

\begin{equation}
F(t) = \frac{2 A}   { 2^{-\{\frac{t-t_{\rm peak}}{t_{\rm doub}}\} } +  2^{+\{{\frac{t-t_{\rm peak}}{t_{\rm half}}} \}   }}  
\label{eq:flareshape}
\end{equation}

\begin{figure*}
  \includegraphics[width=1.98\columnwidth]{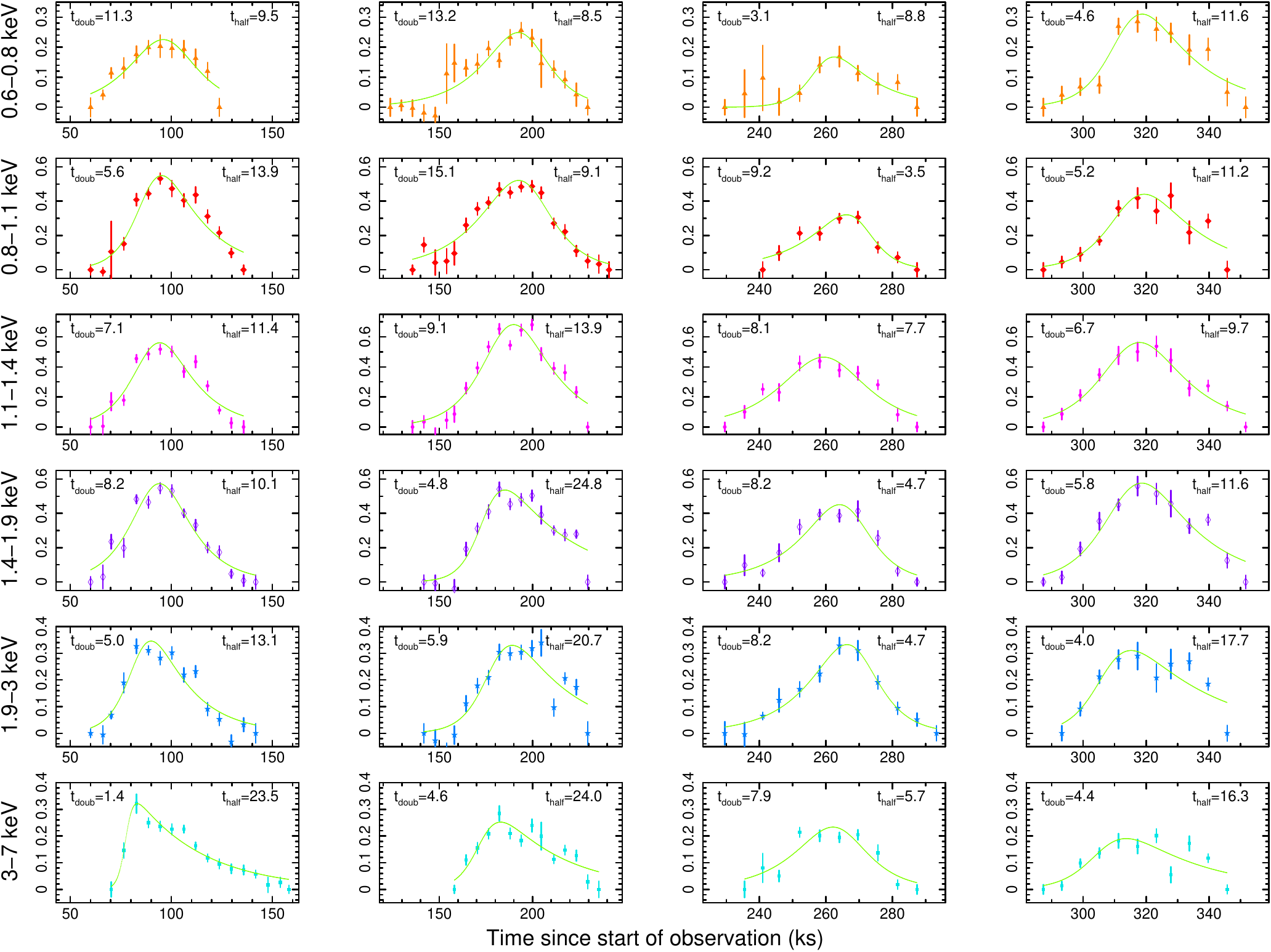}
    \caption{The four individual X-ray flares, fit separately within
      each sub-band following $\S$3.2. The green line denotes the
      best-fitting flare model, and best-fitting values of flux doubling
      ($t_{\rm doub}$) and halving ($t_{\rm half}$) time-scales are listed.  Flares 1 and
      2 (the two left-most columns) have similar rise and decay times
      only towards the softest energies probed; towards harder
      energies, $t_{\rm half}$ increases and $t_{\rm doub}$
      decreases.}
    \label{fig:FLARES}
    \end{figure*}

The four free parameters $t_{\rm peak}$, $t_{\rm doub}$, $t_{\rm half}$, and $A$ denote, respectively, the time of peak flux, the
flux-doubling time-scale, the flux-halving time-scale, and the flare amplitude.
Results are listed in Table~\ref{tab:FLARETABLE}, and the flare fits are plotted in Fig.~\ref{fig:FLARES}.
Uncertainties on one parameter with the other three all free are
always exceptionally large due to strong parameter degeneracies and
relatively small numbers of data points (best-fitting values of the
$\chi^2$ statistic were frequently much less than 1, suggesting
over-parametrization of the data, especially in flares 3 and 4); the
uncertainties on parameters listed in Table~\ref{tab:FLARETABLE} were
therefore determined by holding the other three frozen at their
best-fitting values.
Nonetheless, some trends are apparent: flares 1 and 2 are more
symmetric towards relatively softer energies ($t_{\rm half}$ and
$t_{\rm doub}$ are similar); towards harder energies, $t_{\rm half}$
increases and $t_{\rm doub}$ decreases. Flares 3 and 4 each proceed in
relatively achromatic fashion.

To quantify these trends further, we re-fit each sub-band flare,
assuming a common peak time (the average of the best-fitting values of
$t_{\rm peak}$ across all sub-bands for a given flare).  
In the cases of flares 1 and 2, the values of summed $\chi^2/dof$
(listed in Table~\ref{tab:FLARETABLE}) rose significantly: $F$-tests
indicate that allowing $t_{\rm peak}$ to be a free parameter is
statistically significant at the 92.8 and 98.2\,per cent confidence levels for flares 1
and 2, respectively.  However, for flares 3 and 4, the improvement in
fit is not significant (only 85 and 72\,per cent confidence, respectively).



\begin{table*}
        \centering
        \begin{tabular}{llccccccc} \hline
          \multicolumn{9}{c}{\textbf{Flare Fit Parameters}}  \\ \hline
          Band & Flare start--end  & $N_{\rm pts}$ &  $t_{\rm doub}$  & $t_{\rm half}$ & $t_{\rm peak}$  & $A$  & $\chi^2/dof$ & $\chi^2/dof$ \\
          (keV) & (ks)            &           & (ks)           &    (ks)        &  (ks)        & (ct s$^{-1}$)   &  ($t_{\rm peak}$ free) & ($t_{\rm peak}$ fixed)  \\      \hline
          \multicolumn{9}{c}{Flare 1}  \\ \hline
          {\color{Orange}{0.6--0.8}}    &  60.1--123.8 & 12 & $11.3\pm1.4$     & $ 9.5^{+2.2}_{-1.8}$ & $ 97.2\pm2.3$      & $0.22\pm0.02$  & 3.07/8 & 3.66/9  \\
          {\color{red}{0.8--1.1}}       &  60.1--135.7 & 14 & $ 5.6\pm0.8$     & $13.9\pm1.1$      & $ 90.0\pm1.2$      & $0.50\pm0.03$     & 13.39/10 & 14.80/11 \\
          {\color{Magenta}{1.1--1.4}}   &  60.1--135.7 & 14 & $ 7.1\pm0.9$     & $11.4^{+0.9}_{-0.8}$ & $ 91.2^{+1.2}_{-1.0}$& $0.55\pm0.03$  & 15.47/10  &  17.24/11 \\
          {\color{Purple}{1.4--1.9}}    &  60.1--141.8 & 15 & $ 8.2\pm0.8$     & $10.1\pm0.7$      & $ 92.7\pm1.0$      & $0.57\pm0.03$      & 12.65/11  &  14.78/12 \\
          {\color{RoyalBlue}{1.9--3.0}} &  60.1--141.8 & 15 & $ 5.0\pm0.7$     & $13.1^{+1.1}_{-1.0}$ & $ 85.0\pm1.2$      & $0.31\pm0.02$    & 12.46/11  &  14.04/12 \\
          {\color{ProcessBlue}{3--7}}       &  70.3--158.1 & 16 & $ 1.4^{+0.9}_{-0.7}$& $23.5^{+1.5}_{-1.4}$ & $ 77.7^{+1.0}_{-0.9}$& $0.20\pm0.01$  & 5.82/12  &  10.80/13 \\ 
                                        &              &    &                     &                      &                      &    & {\bf 62.87/62} & {\bf 75.33/68} \\ \hline
          \multicolumn{9}{c}{Flare 2}  \\ \hline 
          {\color{red}{0.6--0.8}}       & 123.8--229.6 & 19 & $13.2^{+1.8}_{-1.6}$& $ 8.5^{+1.8}_{-1.6}$ & $195.6\pm2.4$ & $0.24\pm0.02$    & 10.25/15  &  12.63/16  \\
          {\color{Orange}{0.8--1.1}}    & 135.7--241.0 & 19 & $15.1\pm1.2$     & $ 9.1^{+1.0}_{-0.9}$ & $196.7\pm1.4$ & $0.51\pm0.02$       & 14.48/15  &  21.08/16 \\
          {\color{Magenta}{1.1--1.4}}   & 135.7--229.6 & 17 & $ 9.1\pm0.8$     & $13.9\pm0.9$       & $186.5\pm1.2$ & $0.67\pm0.03$        & 32.51/13  &  35.14/14 \\
          {\color{Purple}{1.4--1.9}}    & 141.8--229.6 & 16 & $ 4.8\pm0.9$     & $24.8\pm1.7$       & $175.7\pm1.3$ & $0.42\pm0.02$      & 25.63/12  &  32.19/13 \\
          {\color{RoyalBlue}{1.9--3.0}} & 141.8--229.6 & 16 & $ 5.9^{+1.2}_{-1.0}$& $20.7^{+2.2}_{-2.0}$ & $180.8\pm1.7$ & $0.28\pm0.02$   & 17.53/12  &  19.39/13\\
          {\color{ProcessBlue}{3--7}}       & 158.1--235.6 & 14 & $ 4.6^{+1.0}_{-0.9}$& $24.0^{+2.4}_{-2.1}$ & $173.7\pm1.6$ & $0.20\pm0.01$   & 8.13/10  &  11.21/11 \\ 
                                        &              &    &                     &                      &                      &     & {\bf 108.52/77} & {\bf 131.65/83} \\ \hline
          \multicolumn{9}{c}{Flare 3}  \\ \hline
          {\color{Orange}{0.6--0.8}}    & 229.6--287.4 & 11 & $ 3.1^{+2.0}_{-1.4}$& $ 8.8^{+2.4}_{-2.1}$ & $259.1^{+2.2}_{-2.3}$& $0.15\pm0.03$   & 1.31/7  &  1.68/8 \\
          {\color{red}{0.8--1.1}}       & 241.0--287.4 &  9 & $ 9.2^{+1.7}_{-1.4}$& $ 3.5^{+1.0}_{-0.9}$ & $269.7\pm1.5$      & $0.29\pm0.03$   & 1.24/5  &  1.94/6 \\
          {\color{Magenta}{1.1--1.4}}   & 229.6--287.4 & 11 & $ 8.1\pm1.1$     & $ 7.7\pm1.0$       & $259.6^{+1.6}_{-1.5}$& $0.46\pm0.04$    & 5.81/7  &  7.06/8 \\
          {\color{Purple}{1.4--1.9}}    & 229.6--287.4 & 11 & $ 8.3^{+0.9}_{-0.8}$& $ 4.7\pm0.7$       & $266.5^{+1.3}_{-1.2}$& $0.43\pm0.04$ & 4.37/7  &  5.08/8 \\
          {\color{RoyalBlue}{1.9--3.0}} & 229.6--293.3 & 12 & $ 8.2\pm0.9$     & $ 4.7^{+0.8}_{-0.7}$  & $268.7\pm1.2$     & $0.32\pm0.03$    & 0.87/8  &  1.65/9 \\
          {\color{ProcessBlue}{3--7}}       & 235.6--287.4 & 10 & $ 7.9^{+1.2}_{-1.1}$& $ 5.7^{+1.0}_{-0.9}$  & $263.7\pm1.4$     & $0.23\pm0.02$ & 3.69/6  &  4.31/7 \\ 
                                        &              &    &                     &                      &                      &     & {\bf 17.30/40} & {\bf 21.71/46} \\ \hline
          \multicolumn{9}{c}{Flare 4}  \\ \hline
          {\color{Orange}{0.6--0.8}}    & 287.4--351.6 & 12 & $ 4.6^{+1.0}_{-0.9}$& $11.6^{+1.9}_{-1.7}$  & $314.4\pm1.6$ & $0.28\pm0.03$  & 3.69/8  &  4.22/9\\
          {\color{red}{0.8--1.1}}       & 287.4--345.7 & 11 & $ 5.2\pm0.8$      & $11.2^{+2.0}_{-1.8}$ & $315.5\pm1.5$  & $0.41\pm0.04$  & 5.69/7  &  6.69/8\\
          {\color{Magenta}{1.1--1.4}}   & 287.4--351.6 & 12 & $ 6.7\pm0.8$      & $ 9.7\pm1.0$       & $315.8\pm1.4$ & $0.55\pm0.04$   & 5.18/8  &  6.21/9 \\
          {\color{Purple}{1.4--1.9}}    & 287.4--351.6 & 12 & $ 5.8\pm0.8$      & $11.6\pm1.1$       & $314.8\pm1.2$ & $0.54\pm0.04$  & 9.13/8  &  10.52/9\\
          {\color{RoyalBlue}{1.9--3.0}} & 293.3--345.7 & 10 & $ 4.0^{+1.1}_{-1.0}$ & $17.7^{+2.7}_{-2.3}$ & $308.2\pm1.5$ &  $0.25\pm0.02$   & 5.30/6  &  6.35/7 \\
          {\color{ProcessBlue}{3--7}}       & 287.4--345.7 & 11 & $ 4.4^{+1.1}_{-0.9}$ & $16.3^{+2.4}_{-2.2}$ & $307.1\pm1.7$ &  $0.16\pm0.02$ & 5.92/7  &  7.03/8 \\ 
                                        &              &    &                     &                      &                      &     & {\bf 34.91/44} & {\bf 41.02/50} \\ \hline 
\end{tabular}
\caption{Best-fitting parameters to individual flares following Eqn.~1. 
Parameter uncertainties correspond to ${\Delta}\chi^2=+2.71$ above the best fit. 
Fits were performed with all four parameters free, yielding the values of $\chi^2/dof$ in Col.~[8].
We then refit, using $t_{\rm peak}$ fixed at the value averaged across
all sub-bands for a given flare, yielding the values of $\chi^2/dof$
in Col.~[9]; this analysis supports the notion that flares peak at earlier times
towards relatively harder X-rays in flares 1 and 2 only.
}
\label{tab:FLARETABLE}
\end{table*}

\section{Cross-Correlations}

\subsection{Intra-X-ray Cross-Correlation Functions}

Previous X-ray lag measurements in Mkn~421 covering both soft and hard
X-rays (e.g., using \textit{ASCA} or \textit{XMM-Newton} and very hard
X-rays (tens of keV; \textit{INTEGRAL}) have yielded a range of
results: some flares yield no non-zero lag detection, and others
sometimes yield positive or negative lags up to a few ks
\citep[e.g.,][]{Takahashi96,Fossati00,Takahashi00,Tanihata01,Ravasio04,Lichti08}. 
Measurements of a hard-to-soft lag, could for example, yield
information on the cooling time-scale, and therefore a limit on the
magnetic field strength, e.g., \citet{Chiappetti99}. 
%

%
We searched for potential lags/leads between the various sub-band
light curves.
We remind the reader that, as pointed out by e.g., \cite{Tanihata01}, 
lags do not necessarily indicate one band "driving" the other
as in e.g., Broad Line Region
reverberation mapping; in blazars, one is quantifying the energy
dependence of decay or rise time-scales of individual flares.
We used both the Discrete Correlation Function
\citep[DCF;][]{Edelson88} and the Interpolated Correlated Function
(ICF; \citealt{White94}; based on \citealt{Gaskell86}).
For the DCF,
we used a bin time of 5.83\,ks (satellite orbtial time
scale); for the ICF; we used a bin time of
2.915\,ks (half an orbit), and we applied the bootstrap error method of
\citet{Peterson98} (both random subset selection and flux
randomization).  We explored lags for the entire duration, as well as
just for each half of the observation.  The results are summarized in
Table~\ref{tab:ccf}. In all cases, a positive lag indicates that the
relatively harder band leads the softer band.
We find evidence for the 0.6--0.8\,keV band to lead the other bands by
an average of $4.6\pm2.6$\,ks (from ICF centroids), but only during the
first half of the observation.  For other waveband combinations, no
lags are detected.

From Fig.~\ref{fig:SXTlcoverplot}, the softest band visually leads the
other bands mostly at local minima, i.e., flares 1 and 2 seem to start
a bit earlier in the 0.6--0.8\,keV band, consistent with this finding.
However, measured lag values are the net sum of all lags/leads within
the light curve pair.  From the fits to individual flares above,
relatively harder bands peak earlier than softer bands during flares 1
and 2.  Both positive and negative lag-generating processes are
concurrently at work.

When performing cross correlations on pairs of light curves that suffer
from recurring gaps, one must apply caution and confirm that any
lags/leads observed have not arisen spuriously as an artefact of the
sampling -- particularly if the observed lags/leads are on the same
order as the periodic gaps \citep[e.g.,][]{Takahashi00, Edelson01,
Sembay02}.  Observations using satellites in low-Earth orbit, such as
\textit{AstroSat}, can have periodic gaps due to Earth occultations. \textit{AstroSat}'s orbital period is 5.83\,ks,
a value not far from the $4.6\pm2.6$\,ks lag observed from the 0.6--0.8\,keV band to the
higher-energy bands (we adopt the average of the lags from the
0.6--0.8\,keV band to each of the other bands) during the first half of the observation.
The fact that only the softest light curve shows a lead while all bands
are impacted equally by data gaps might support its divergence from
the other bands as being real, but we nonetheless conducted such
simulations to assess the impact on the measured lag value.

Here, we perform Monte Carlo simulations, starting with the null hypothesis that
the two light curves are intrinsically correlated with zero lag.  With
the algorithm of \citet{TK95}, we simulated
500 pairs of light curves, using the same random number seed for each. 
%
The input PSD slope was an unbroken power law of slope $-$2.2 for both
light curves, consistent with \citet{Chatterjee18}; the energy
dependence of PSD shape within the X-ray band has, to our knowledge,
not been explored thoroughly yet so we explore a wider range of slopes
below.
PSD amplitudes were chosen such that the integral of the PSDs from
1/(194\,ks)=$5.2\times10^{-6}$\,Hz to the Nyqust frequency of 1/(2$\times$5.3\,ks) = $8.6\times10^{-5}$\,Hz
matched the measured values of $F_{\rm var}$ for the first half of the 0.6--0.8\,keV (hereafter ``S'')
and 1.4--1.9\,keV (hereafter ``H'') light curves, 7.4 and 10.8\,per cent, respectively.
We simulated light curves with
a time resolution of 100\,s (to match the SXT extraction) and a duration of 940\,ks, (to allow
for red noise leakage), trimmed them to 188\,ks, and orbitally binned
them in the exact same manner as for the observed light curves.
We re-scaled each simulated light curve to match the mean and standard deviation of the S and H light curves.
We added Poisson noise by deviating each binned data point by a Gaussian
whose standard deviation is equal to the average count rate error
within each bin in the observed light curves (0.034 and 0.040\,ct~s$^{-1}$ for S and H, respectively).
We measured each ICF centroid,
and examined the distribution of ICF centroids: only 13/500 trials
yields soft-to-hard lags greater than the observed lag lower limit of 2.0\,ks
(rejection of null hypothesis at 97.4\,per cent confidence).

Finally, we explored the potential impact of differing PSD power-law
slopes, considering the best-fitting power-law slope values as discussed in Appendix~A.
We re-simulated the light curve pairs (200 trials), considering power-law slopes
of $-$2.0 for both light curves, $-$2.4 for both light curves, $-$2.0 for
soft and $-$2.4 for hard, and $-$2.4 for soft and $-$2.0 for hard, each
time keeping the PSD normalized to match the observed values of
$F_{\rm var}$, and again including Poisson noise.  Again adopting a common observed soft-to-hard lag
limit of +2.0\,ks, we find that for these four cases, we can reject the null 
hypothesis at 98.5, 95.0, 98.5, and 95.0\,per cent 
confidence, respectively.

Note, however, that all these simulations do not account for the fact that we observed a lag in
the first half of the data only; the second half of the data, in which we observed no lag,
effectively provides a doubling in the number of trials. To account for this
aspect of a "look elsewhere effect" \citep{Algeri16}, we 
double the likeliood of the null hypothesis being correct:
$94.8$, $90.0$, $97.0$\,per cent, (PSD slopes of $-$2.2, $-$2.0, $-$2.4 for both bands, respectively),  
$97.0$ and $90.0$\,per cent ($-$2.0 for S and $-$2.4 for H;  $-$2.4 for S and $-$2.0 for H, respectively).  

%
%
We conclude that the observed soft-to-hard lag is likely intrinsic to the source
and is not an artefact of data sampling or Poisson noise.


\begin{table*}
        \centering
        \begin{tabular}{llcccc} \hline
	  \multicolumn{6}{c}{\textbf{Intra-X-ray Cross Correlations}}  \\ \hline
          \multicolumn{6}{c}{\textit{Full Duration}}  \\ \hline
 first        & second       & DCF $r_{\rm corr}$   & DCF $\tau$ (ks) & ICF centroid $r_{\rm corr}$   & ICF centroid $\tau$ (ks) \\
 LAX 3--7     & {\color{ProcessBlue}{SXT 3--7}}     & 0.9219$\pm$0.0875  &     0           &  0.9296  &   --0.774$\pm$1.338  \\ 
 {\color{ProcessBlue}{SXT 3--7}}    & {\color{RoyalBlue}{SXT 1.9--3}}   & 0.9403$\pm$0.1128  &     0           &  0.9403  &    +0.340$\pm$1.406  \\ 
 {\color{ProcessBlue}{SXT 3--7}}     & {\color{Purple}{SXT 1.4--1.9}} & 0.9220$\pm$0.0947  &     0           &  0.9214  &   --0.085$\pm$1.562   \\ 
 {\color{ProcessBlue}{SXT 3--7}}     & {\color{Magenta}{SXT 1.1--1.4}} & 0.9044$\pm$0.0985  &     0           &  0.9040  &   --0.468$\pm$1.592  \\ 
 {\color{ProcessBlue}{SXT 3--7}}     & {\color{red}{SXT 0.8--1.1}} & 0.8794$\pm$0.1059  &     0           &  0.8951  &    +0.650$\pm$2.088   \\  
 {\color{ProcessBlue}{SXT 3--7}}     & {\color{Orange}{SXT 0.6--0.8}} & 0.8014$\pm$0.1068  &     0           &  0.8282  &  --1.621$\pm$2.622  \\  \hline

 {\color{ProcessBlue}{SXT 3--7}}     & {\color{Orange}{SXT 0.6--0.8}} & 0.8014$\pm$0.1068  &     0           &  0.8282  &  --1.621$\pm$2.622  \\ 
 {\color{RoyalBlue}{SXT 1.9--3}}   & {\color{Orange}{SXT 0.6--0.8}} & 0.8874$\pm$0.1022  &     0           &  0.9090  &  --1.620$\pm$2.169   \\ 
 {\color{Purple}{SXT 1.4--1.9}} & {\color{Orange}{SXT 0.6--0.8}} & 0.9198$\pm$0.0847  &     0           &  0.9320  &  --1.065$\pm$1.700   \\ 
 {\color{Magenta}{SXT 1.1--1.4}} & {\color{Orange}{SXT 0.6--0.8}} & 0.9454$\pm$0.0801  &     0           &  0.9511  &  --1.030$\pm$2.154   \\ 
 {\color{red}{SXT 0.8--1.1}} & {\color{Orange}{SXT 0.6--0.8}} & 0.9417$\pm$0.0906  &     0           &  0.9598  &  {\bf --2.337$\pm$2.100}  \\  \hline
          \multicolumn{6}{c}{\textit{First Half (0--194\,ks)}}  \\ \hline
 first        & second       &  DCF $r_{\rm corr}$    & DCF $\tau$ (ks) & ICF centroid $r_{\rm corr}$   & ICF centroid $\tau$ (ks) \\
 LAX 3--7     & {\color{ProcessBlue}{SXT 3--7}}     & 0.8892 $\pm$ 0.1398  &  0             &  0.8990 &   --0.873$\pm$1.675   \\ 
 {\color{ProcessBlue}{SXT 3--7}}     & {\color{RoyalBlue}{SXT 1.9--3}}   & 0.9362 $\pm$ 0.0116  & 0              &  0.9464 &   --0.834$\pm$1.711     \\ 
 {\color{ProcessBlue}{SXT 3--7}}     & {\color{Purple}{SXT 1.4--1.9}} & 0.8740 $\pm$ 0.1248  & 0              &  0.8963 &  --1.193$\pm$2.003    \\ 
 {\color{ProcessBlue}{SXT 3--7}}     & {\color{Magenta}{SXT 1.1--1.4}} & 0.8544 $\pm$ 0.0898  & 0              &  0.8809 &  --1.835$\pm$1.891    \\ 
 {\color{ProcessBlue}{SXT 3--7}}     & {\color{red}{SXT 0.8--1.1}} & 0.7266 $\pm$ 0.1658  & --5.8          &  0.8503 &  --1.009$\pm$2.899  \\ 
 {\color{ProcessBlue}{SXT 3--7}}     & {\color{Orange}{SXT 0.6--0.8}} & 0.7631 $\pm$ 0.1468  & --5.8          &  0.7648 &  {\bf --5.524$\pm$3.206}   \\   \hline

 {\color{ProcessBlue}{SXT 3--7}}     & {\color{Orange}{SXT 0.6--0.8}} & 0.7631 $\pm$ 0.1468  & --5.8          &  0.7648 &  {\bf --5.524$\pm$3.206}  \\ 
 {\color{RoyalBlue}{SXT 1.9--3}}   & {\color{Orange}{SXT 0.6--0.8}} & 0.8547 $\pm$ 0.1626  & --5.8          &  0.8551 &  {\bf --5.044$\pm$2.689}    \\   
 {\color{Purple}{SXT 1.4--1.9}} & {\color{Orange}{SXT 0.6--0.8}} & 0.8776 $\pm$ 0.1249  & --5.8          &  0.8816 &  {\bf --4.366$\pm$2.314}  \\ 
 {\color{Magenta}{SXT 1.1--1.4}} & {\color{Orange}{SXT 0.6--0.8}} & 0.9099 $\pm$ 0.1541  & --5.8          &  0.9344 &  {\bf --4.004$\pm$2.333}    \\  
 {\color{red}{SXT 0.8--1.1}} & {\color{Orange}{SXT 0.6--0.8}} & 0.9145 $\pm$ 0.1667  & --5.8          &  0.9329 &  {\bf --3.926$\pm$2.426}   \\  \hline
 \multicolumn{6}{c}{\textit{Second Half (194--389\,ks)}}  \\ \hline
 first        & second       &  DCF $r_{\rm corr}$   & DCF $\tau$ (ks) & ICF centroid $r_{\rm corr}$   & ICF centroid $\tau$ (ks) \\
 LAX 3--7     & {\color{ProcessBlue}{SXT 3--7}}     & 0.8657 $\pm$ 0.1515 &  0              &  0.8925 &    +0.116$\pm$1.713  \\ 
 {\color{ProcessBlue}{SXT 3--7}}     & {\color{RoyalBlue}{SXT 1.9--3}}   & 0.8658 $\pm$ 0.1170 &  0              &  0.8690 &    +0.856$\pm$2.163  \\ 
 {\color{ProcessBlue}{SXT 3--7}}     & {\color{Purple}{SXT 1.4--1.9}} & 0.8626 $\pm$ 0.1019 &  0              &  0.8806 &    +0.283$\pm$1.971   \\ 
 {\color{ProcessBlue}{SXT 3--7}}     & {\color{Magenta}{SXT 1.1--1.4}} & 0.8179 $\pm$ 0.1016 &  0              &  0.8219 &   --0.268$\pm$1.950  \\ 
 {\color{ProcessBlue}{SXT 3--7}}     & {\color{red}{SXT 0.8--1.1}} & 0.7424 $\pm$ 0.1376 &  0              &  0.8172 &   +1.357$\pm$1.840   \\ 
 {\color{ProcessBlue}{SXT 3--7}}     & {\color{Orange}{SXT 0.6--0.8}} & 0.6754 $\pm$ 0.1456 &  0              &  0.7229 &   +1.867$\pm$2.498   \\ \hline
 {\color{ProcessBlue}{SXT 3--7}}     & {\color{Orange}{SXT 0.6--0.8}} & 0.6754 $\pm$ 0.1456 &  0              &  0.7229 &   +1.867$\pm$2.498    \\ 
 {\color{RoyalBlue}{SXT 1.9--3}}   & {\color{Orange}{SXT 0.6--0.8}} & 0.7855 $\pm$ 0.1018 &  0              &  0.7890 &    +0.992$\pm$2.350  \\ 
 {\color{Purple}{SXT 1.4--1.9}} & {\color{Orange}{SXT 0.6--0.8}} & 0.8722 $\pm$ 0.1172 &  0              &  0.8740 &   +1.018$\pm$2.661   \\ 
 {\color{Magenta}{SXT 1.1--1.4}} & {\color{Orange}{SXT 0.6--0.8}} & 0.8886 $\pm$ 0.1177 &  0              &  0.8904 &   +1.295$\pm$2.357  \\ 
 {\color{red}{SXT 0.8--1.1}} & {\color{Orange}{SXT 0.6--0.8}} & 0.8915 $\pm$ 0.1025 &  0              &  0.9052 &  --1.111$\pm$2.478   \\  \hline
\end{tabular}
        \caption{DCF and ICF centroid results.
          A positive lag indicates the first band leading the second band. Lags not consistent with zero are in bold font.
          Results indicate that the softest band leads the other bands during the first half of
          the observation. 
}
\label{tab:ccf}
\end{table*}


\subsection{X-ray/TeV Correlated Behaviour}\label{sec:xtzerolag}



\begin{figure}
  \includegraphics[width=0.95\columnwidth]{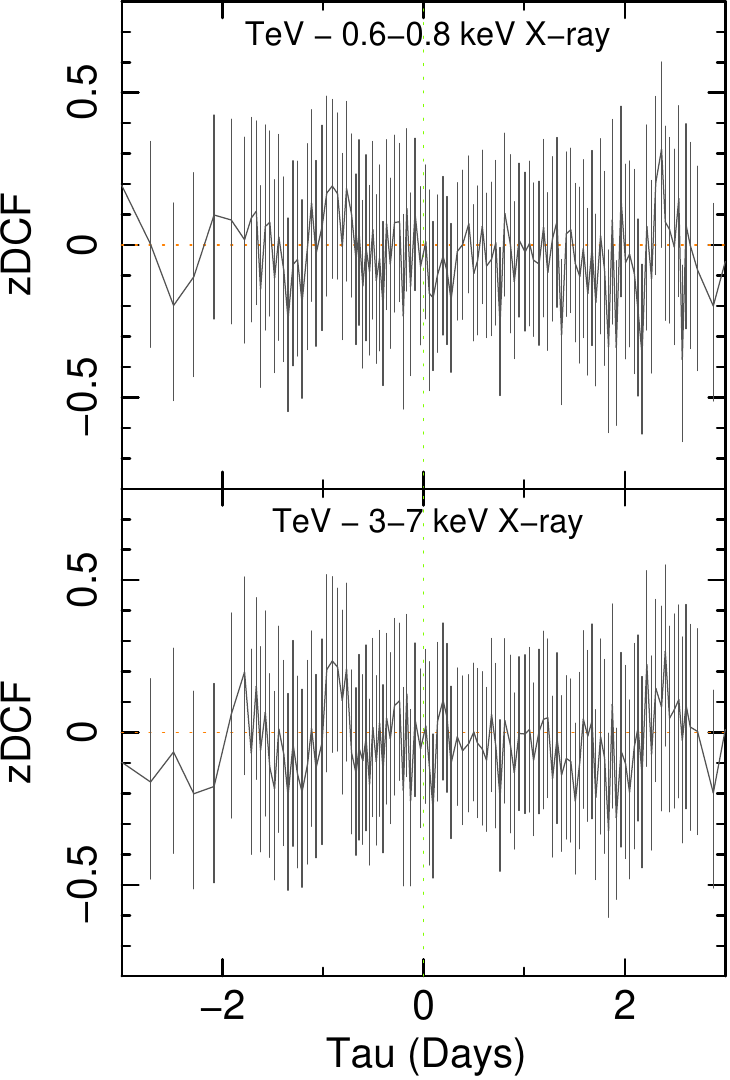}    
  \caption{Cross-correlation results for the TeV band (40-minute bins)
    versus the orbitally-binned soft (0.6--0.8\,keV) and hard (3--7\,keV) bands.  A
    positive lag indicates TeV leading the X-ray band.}
    \label{fig:plotzdcfs}
    \end{figure}

We first searched for lags between the 40-minute binned TeV band and the
softest and hardest X-ray bands (0.6--0.8 and 3--7 keV); given the
data daps in the TeV light curve, we used the z-transformed Discrete
Correlation Function \citep[zDCF;][]{Alexander97,Alexander13}.
We do not find any robust evidence for a correlation between the
TeV band and either X-ray band, with maximum correlation coefficients
remaining below 0.3 on time scales between $\pm2$\,days
in both cases.
Both zDCFs are plotted in Fig.~\ref{fig:plotzdcfs}.
 %
 %
We additionally explored zero-lag correlations by
plotting TeV fluxes as a function of X-ray flux (either soft
or hard band), using data points as close as possible to each other in
time. Neither across all four nights nor within any one night (where very
small-number statistics dominate) do we see any evidence for a
correlation. 
   
In conclusion, we cannot claim any TeV--X-ray correlation, in part due
to the fact that our campaign simply did not sample strong inter-night
variations in either band, and also due to the
modest flux level of the source yielding
relatively large flux uncertainties within each time bin.
Consequently, constraints on whether TeV
flux varies proportionally to X-ray flux e.g., in a linear or
quadratic fashion are not attainable from these data.

\section{X-ray spectral modeling}

We first modeled the time-averaged 0.6--7\,keV SXT and 3--7\,keV LAXPC spectrum.
We used Interactive Spectral Interpretation System \textsc{(ISIS)}
v.1.6.2-44 for all spectral fitting.  We grouped the SXT data to a
minimum signal/noise of 5 within \textsc{ISIS}.  For all fits, we
applied an instrumental cross-normalization constant, frozen to 1 for
SXT and allowed to be free for LAXPC. To achieve reasonable fits, we
added 2\,per cent systematics to both the time-averaged and
time-resolved LAXPC spectra.

All spectral fits included Galactic absorption accounting for monatomic and molecular hydrogen,
$N_{\rm H, total} = 2.03 \times 10^{20}$\,cm$^{-2}$ 
\citep{Willingale13xx},\footnote{\url{https://www.swift.ac.uk/analysis/nhtot/index.php}},
and used a redshift value of $z=0.030021$ \citep{DeVaucouleurs91}.

\subsection{Time-averaged X-ray spectral modeling}

All of our model fits to the time-average spectrum yielded large
residuals and high values of $\chi^2_{\rm red}$, likely due to strong
spectral variability during the observation.  Our time-averaged fits thus serve only as an
approximate guide for forming an initial model to be applied to the
time-resolved spectra in the next subsection.

We first tested a single power law, which yielded $\chi^2/dof = 1651.68/583 = 2.833$.
Our best-fitting model was a broken power law ($\chi^2/dof$ = 1590.08/581 = 
2.737), with photon index steepening from $\Gamma_1$ = $2.200\pm0.005$
below a break energy of $5.0^{+0.3}_{-0.2}$\,keV to $\Gamma_2$ =
$2.726^{+0.218}_{-0.134}$ above it, and indicating some slow spectral
curvature across the bandpass.

We use the Akaike Information Criterion \citep[AIC;][]{Akaike73} 
with finite sample correction \citep{Sugiura78} to
ascertain the significance in improvement: For each model, we
calculate $AIC = 2m - 2C_{\rm L} + \chi^2 + (2m(m+1))/(n-m-1)$,
where $m$ is the number of free parameters (5 for unbroken, 7 for broken), $n$ is the number of data
bins (580 for SXT + 8 for LAXPC), and $C_{\rm L}$ is the likelihood of the true model. We compare the models
by computing the difference in values of $AIC$, so the $C_{\rm L}$ terms cancel;
${\Delta}AIC = -57.5$, indicating that the broken power-law model is a better fit.
We caution, though, that while preference for this model supports
  the presence of some spectral curvature, the magnitude of this
  (phenomenological) spectral break is very strong compared to the
  best-fitting physically-motivated SED models, discussed in
  $\S$\ref{sec:sed}. That is, we caution against interpreting the
  inferred change in slope ($\sim0.5$) too literally.

\begin{figure}
  \includegraphics[width=0.99\columnwidth]{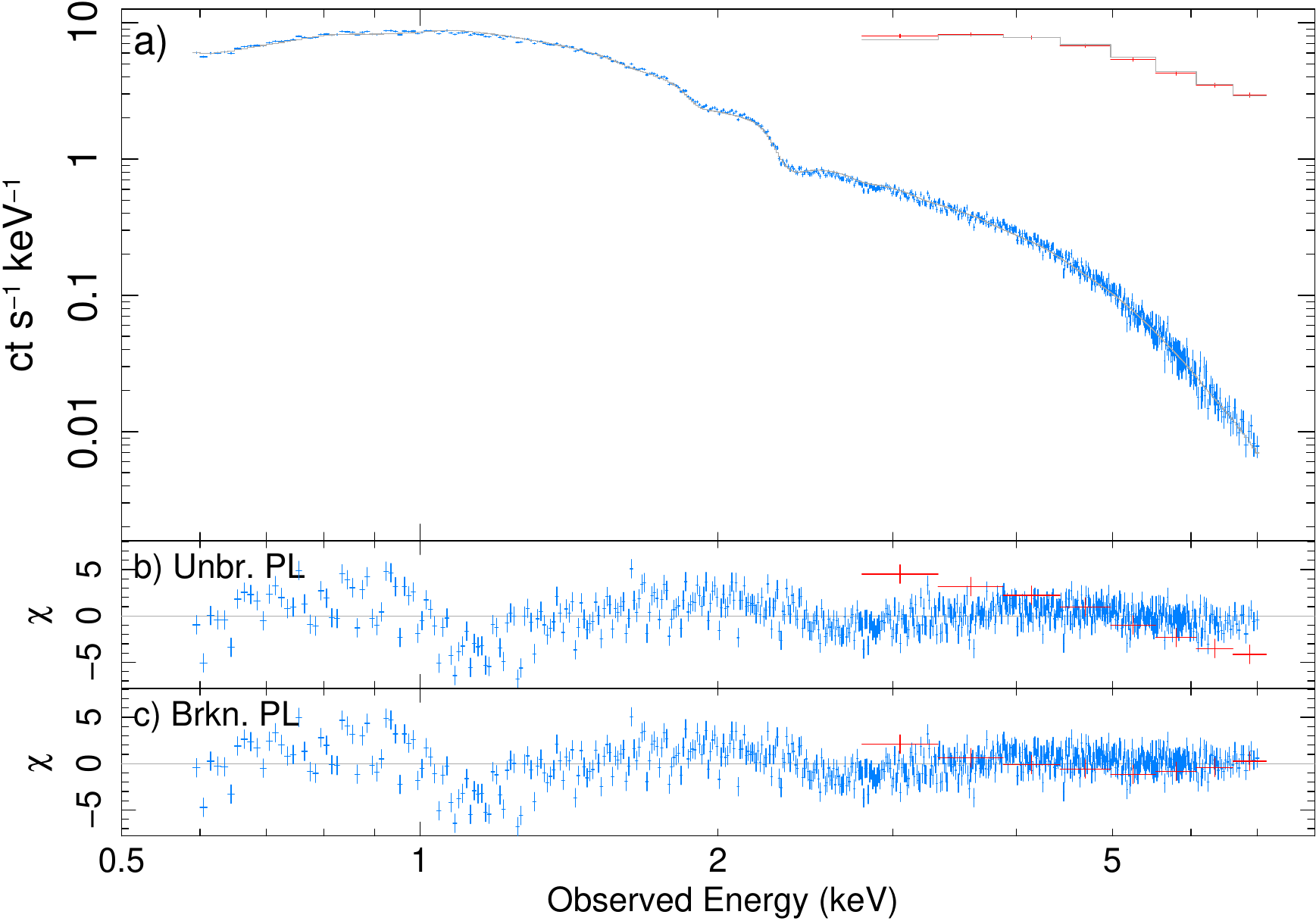}
  \caption{Time-averaged spectral counts data are plotted in panel a).
SXT and LAXPC data are shown in blue and red, respectively; the gray lines denote the best-fitting broken power-law model.
Panels b) and c) show the residuals to our best-fitting
    unbroken and broken power-law fits, respectively;
    most of the improvement in the broken power-law fit compared to the unbroken power-law fit
    is at higher energies.}
    \label{fig:TAfitsdatares}
    \end{figure}

Finally, we applied a phenomenological log-parabola fit to model this curvature:
$A(E) = K ( E(1+z)/E_{\rm pivot} )^{(-a - b \textrm{log}( E(1+z)/E_{\rm pivot}))}$;
it does not contain a sharp break like the broken power law.
We froze $E_{\rm pivot}$ at 0.7\,keV.
However, our best-fitting model yielded $b$ consistent with zero, a value of $a$ identical to that of $\Gamma$ in the single power-law fit,
and an identical value of $\chi^2_{\rm red}$ as well, so we do not discuss log-parabola fits further.
The $\chi$ residuals for the unbroken and broken power-law models are presented in Fig.~\ref{fig:TAfitsdatares},
and the best-fitting model parameters are listed in Table~\ref{tab:TAfits}.

\begin{table*}
        \centering
        \begin{tabular}{lcc} \hline
	\multicolumn{3}{c}{\textbf{Fits to the Time-Averaged SXT + LAXPC Spectrum}}  \\ \hline
Parameter              &  Single Power Law                  &    Broken Power Law   \\
$\chi^2$/dof           &  1651.68/583=2.833                & 1590.08/581=2.737      \\
$\Gamma$               &  $2.204\pm0.004$                   &     ---                              \\
$\Gamma_1$             &             ---                    & $2.20\pm0.01$     \\
$\Gamma_2$             &             ---                    & $2.726^{+0.22}_{-0.13}$     \\
$E_{\rm break}$ (keV)  &             ---                    & $5.0^{+0.3}_{-0.2}$               \\
Normalization          &  $0.184\pm0.001$                   & $0.184\pm0.001$     \\ 
SXT Gain Slope         &  $1.0*$                            & $1.0*$               \\
SXT Gain Intercept     &  $0.034^{+0.001}_{-0.002}$         & $0.033^{+0.001}_{-0.002}$     \\
LAX Constant           &  $0.88\pm0.02$                     & $0.92\pm0.02$          \\
LAX Gain Slope         &  $1.0*$                            & $1.0*$                 \\
LAX Gain Intercept     &  $-0.55^{+0.05}_{-0.10}$ &  $-0.37\pm0.09$  \\
0.6--6\,keV model flux  & $5.828\times10^{-10}$ (17.28 mCrb) & $5.824\times10^{-10}$ (17.26 mCrb)  \\   \hline 
        \end{tabular}
\caption{Normalization refers to the 1\,keV normalization (ph~cm$^{-2}$~s$^{-1}$~keV$^{-1}$). 
Model flux here refers to observed/absorbed flux in units of erg~cm$^{-2}$~s$^{-1}$.
An asterisk (*) indicates a frozen parameter.}
\label{tab:TAfits}
\end{table*}

\subsection{Time-resolved X-ray spectral modeling}

We fit only those orbits containing both SXT and LAXPC data, and thus
excluded the first orbit (which had LAXPC only) and the last three
orbits (SXT only).  We first fit a single power-law model to each of
the 63 spectra; the average value of $\chi^2_{\rm red}$ was 1.20.  We
then applied a broken power-law model to each spectrum, with
$\Gamma_1$, $\Gamma_2$, and $E_{\rm break}$ free but initially set to
their best-fitting values from the time-average spectrum.  This model
yielded a superior fit: the average value of $\chi^2_{\rm red}$ fell
to 1.05.  $F$-tests performed on each pair of models for each of the
63 orbits always yielded values of $F$ greater than 5.3 and null
hypothesis probabilities less than $0.7$\,per cent (the more complex
model always provides a superior fit at the $\geq 99.3$\,per cent
confidence level) for each orbit.
%
%
%
%
We then re-fit the broken power-law model with $E_{\rm break}$ frozen
to 4.4\,keV, the mean of the values obtained from the time-resolved fits; the
value of $\chi^2/dof$ summed across all orbits rose only from 14766.71/13997  to 14801.48/14060, and an
$F$-test indicated that it was not statistically significant to leave this parameter
thawed across the sample.
%
%
%
%
%

\begin{figure*}
  \includegraphics[width=1.8\columnwidth]{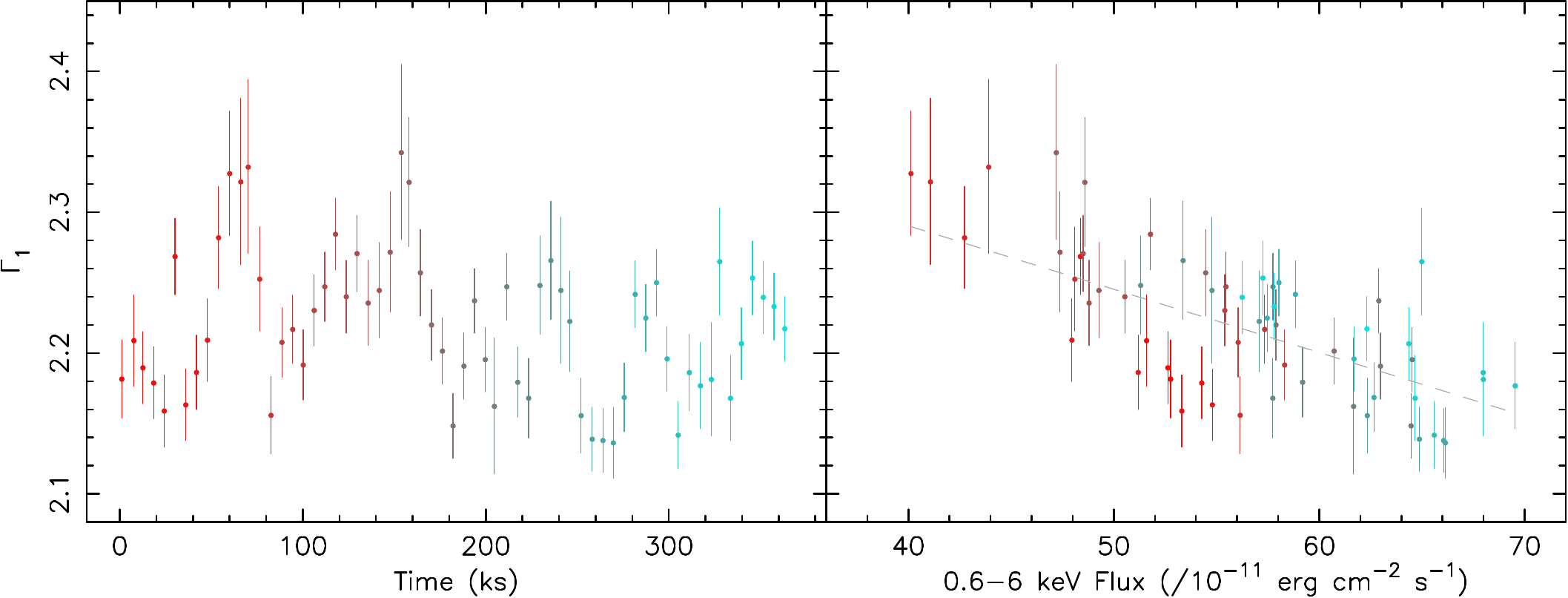}
  \caption{Results from the time-resolved spectral fits using the broken power-law model, with
    $E_{\rm break}$ frozen at 4.4\,keV and $\Gamma_2$ fixed to 
    $\Gamma_1 + 0.49$, where $\Gamma_1$ and $\Gamma_2$ denote the photon indices below and above the break energy, respectively.
  The left and right panels show $\Gamma_1$ as a function of time and flux, respectively; levels of red/cyan colour
  indicate points closer to the beginning/end of the observation.
  Time zero is the same as in Fig.~1.
    The dashed grey line is the best-fitting linear relation. }    
   \label{fig:TRfit_BKN_jan21}
  \end{figure*}

Refitting the time-resolved spectra with $\Gamma_2$ fixed to $\Gamma_1
+ 0.49$ (the mean values of $\Gamma_1$ and $\Gamma_2$ in the time-resolved fits above 
were 2.23 and 2.72, respectively) does
not yield a significantly worse set of fits: the summed $\chi^2$/dof
rose from 14801.48/14060 to 14870.95/14123, yielding an $F$-test null hypothesis
probability of 0.63.  We thus display the results from fits with $\Gamma_2$ fixed to $\Gamma_1 + 0.49$ 
in Fig.~\ref{fig:TRfit_BKN_jan21}.  The best-fitting linear regression to
the $\Gamma_1$ versus 0.6--6.0\,keV model flux relation follows     $\Gamma_1 = (-0.0045\pm0.0006)F_{0.6-6} + (2.47\pm0.03)$.
We also re-plotted the time-resolved spectral fit results of photon
index versus time, dividing up the data into four separate flares
defined by the minima in the 1.42--1.90\,keV light curve at
[60--135], [135--229], [229--286], and [286--353]\,ks to check for any
divergence in their spectral behaviour. However, the spectral and
spectral variability behaviour across all flares overlap well.





We tested for any signs of hysteresis in the $\Gamma$--flux plane, as
has been found for various blazars including Mkn~421
\citep[e.g.,][]{Takahashi96,Cui04,Abeysekara17}, by comparing values
during the rising and decaying flux trends; hysteresis could
potentially provide additional insight into cooling and acceleration
time-scales.  However, no obvious trends were found.  Perhaps the range
in flux sampled by our campaign was simply too small. Alternately and
speculatively, perhaps there are simultaneous contributions from
multiple flares, each in different parts of their hysteresis cycles.

\section{TeV spectral modeling}  

\begin{figure}
  \includegraphics[width=0.95\columnwidth]{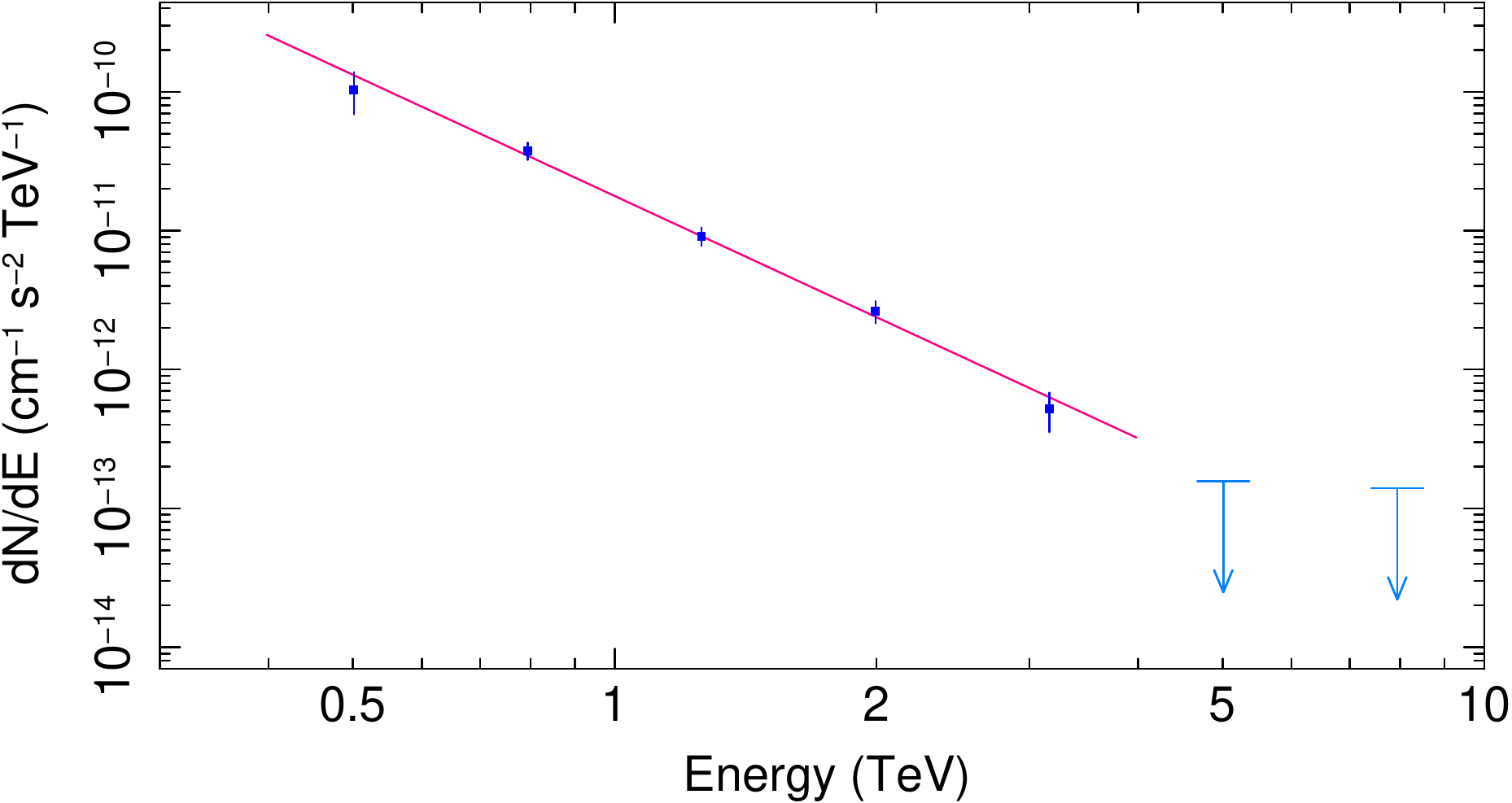}
\caption{FACT TeV spectrum covering 2019 January 10--14.
  The red line is the best least-squares fit to the 0.4--4.0 TeV bins.
These data were not corrected for EBL extinction.}
\label{fig:factcompletespec}
\end{figure}

We present here the spectrum for the complete campaign as measured by FACT,
integrated over 10--14 January.
We extracted fluxes in bins above 0.4 TeV 
in bin widths of 0.2 in the log,
though only upper limits for bins $>4$ TeV were obtained.
The spectrum is plotted in Fig.~\ref{fig:factcompletespec}.
A least-squares fit to the bins from 0.4 to 4 TeV yields a spectral slope
of $-2.90\pm0.16$ and a monochromatic 1 TeV normalization of
$1.78\pm0.16\times10^{-11}$ ph cm$^{-2}$ s$^{-1}$ TeV$^{-1}$.
The $>$0.4 TeV integrated fluxes were 
$4.64\pm0.42 \times 10^{-11}$  ph~cm$^{-2}$~s$^{-1}$,
$3.72\pm0.33 \times 10^{-11}$ TeV~cm$^{-2}$~s$^{-1}$, and
$5.83\pm0.52 \times 10^{-11}$ erg~cm$^{-2}$~s$^{-1}$
(all values here are for data not yet corrected for extragalactic background light extinction).
We also extracted spectra for individual nightly bins to check for
inter-night variability but found no significant evidence for it,
so these nightly spectra are not discussed further.

Finally, in preparation for incorporating the FACT data on to the SED,
we corrected ${\nu}F_{\nu}$ data points for extragalactic background
light (EBL) extinction; we followed the prescription of
\citet{Franceschini08}, which increased flux density at 0.5 and 3.2
TeV (the midpoints of the lowest- and highest-energy bins with
detections) by 14 and 60\,per cent respectively.

\section{SED construction and modeling}\label{sec:sed}

\begin{figure*}
\includegraphics[width=\textwidth]{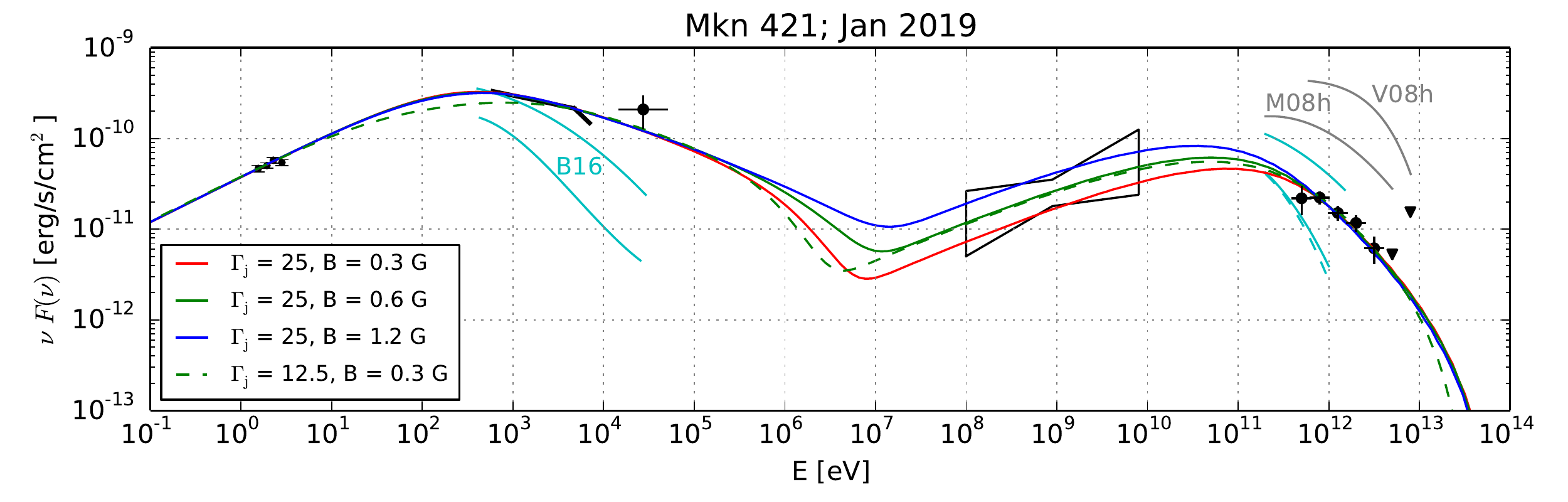}
\caption{Spectral energy distribution (SED) of Mkn~421 during our
  campaign. The results of WEBT, \textit{AstroSat}, and FACT observations in January 2019 are shown in
  black symbols; also included are a concurrent \textit{Fermi}-LAT spectrum and a \textit{Swift}-BAT flux point.
  For comparison, the X-ray and VHE measurements of
  Mkn~421 in two flux states selected from \citet[][their Fig.~13, MJD~56302 and 56335]{Balokovic16} are shown with
  cyan lines.
  The gray lines show the high VHE states measured by
  MAGIC and VERITAS in their 2008 campaigns \citep{Aleksic12,Acciari11}. The broad-band lines show
  one-zone synchrotron+SSC leptonic models matched to the optical, X-ray
  and VHE data.}
\label{fig:fig_sed}
\end{figure*}

We form the SED by combining the time-averaged $B$, $V$, $R$, and $I$ flux
densities, the best-fitting broken power-law fit to the time-averaged
X-ray spectrum, the best fit to the time-averaged TeV spectrum, and
the best-fitting power-law model to the {\it Fermi}-LAT spectrum.  We also
included the concurrent \textit{Fermi}-LAT spectrum from the four-day
period MJD 58493--7: we applied a power-law fit over the 0.1--8 GeV
range, finding $\Gamma = 1.64 \pm 0.22$ and an integrated photon flux
density of $1.11 \pm 0.36 \times 10^{-7}$ ph cm$^{-2}$ s$^{-1}$.
We also included 15--50~keV data from \textit{Swift}'s Burst Alert Transient
(BAT) Hard X-ray Transient Monitor program\footnote{\url{https://swift.gsfc.nasa.gov/results/transients/}} \citep{Krimm13}
for a 30-day period centered on our campaign. One-day-binned count rate light curves were downloaded,
binned to 30\,days, and converted to flux based on the spectrum obtained in the 105 month BAT
survey\footnote{\url{https://swift.gsfc.nasa.gov/results/bs105mon/}}
\citep{Oh18}: $F_{15-50} = 2.09\pm0.80\times10^{-10}$ erg cm$^{-2}$ s$^{-1}$. 
The resulting SED is plotted in Fig.~\ref{fig:fig_sed}.



The SED of Mkn~421 has been fitted with
a family of one-zone leptonic models calculated with the {\tt BLAZAR}
code \citep{Moderski03}.  The code solves a
kinetic equation for the energy distribution of electrons, including
injection of a broken power-law distribution, as well as adiabatic and
radiative losses (including synchrotron and synchrotron self Compton
(SSC) mechanisms) along a conical jet.  In Fig.~\ref{fig:fig_sed} and
Table \ref{tab:tab_sed_params}, we present four SED models that differ by
two key parameters: the jet bulk Lorentz factor $\Gamma_{\rm j}$
and the co-moving magnetic
field strength $B$.
These models have been fitted tightly to the X-ray spectrum observed
by the \textit{AstroSAT} and to the VHE gamma ray spectrum observed by the
FACT, they have also been constrained by the optical flux density and
colour.  The injected electron energy distribution
has been modeled as a broken power-law with two power-law sections:
$N(\gamma) \propto \gamma^{-p_1}$ for $1 < \gamma < \gamma_{\rm br}$,
and $N(\gamma) \propto \gamma^{-p_2}$ for $\gamma_{\rm br} < \gamma <
\gamma_{\rm max}$.  We fixed the low-energy index at $p_1 = 2$, which
determines the hardness of the synchrotron component in the optical
band and of the SSC component in the HE gamma ray (\textit{Fermi}) band.  The
high-energy index $p_2$ has been tuned for each model to match the
\textit{AstroSAT} and FACT spectra.  The value of $\gamma_{\rm br}$ determines
a proper connection of the optical and X-ray data by the synchrotron
component.  The VHE flux level is determined mainly by the
distance $r$ of the emitting region from the jet base, which in turn
determines the lateral jet radius $R_{\rm j} = r\theta_{\rm j}$, where
$\theta_{\rm j} = 0.2/\Gamma_{\rm j}$ is the jet half-opening angle
(assuming a conical geometry).  A larger emitting region results in a
lower Compton dominance $L_{\rm SSC} / L_{\rm syn}$.

We have considered two values of the bulk Lorentz factor motivated by previous studies: $\Gamma_{\rm
  j} = 25$ and $\Gamma_{\rm j} = 12.5$.  Given a value of $\Gamma_{\rm
  j}$, the co-moving magnetic field strength $B$ determines the level of HE
gamma-ray emission (which scales roughly as $\propto {B}^{2/3}$),
which is poorly constrained by the \textit{Fermi}/LAT measurement simultaneous
with the \textit{AstroSAT} campaign.  Optimum levels of HE flux were obtained for $B =
0.6\;{\rm G}$ with $\Gamma_{\rm j} = 25$, and for $B = 0.3\;{\rm G}$
with $\Gamma_{\rm j} = 12.5$.
  Mild cooling breaks manifest themselves in these SED models; analytical
  estimates of cooling breaks can be obtained as follows:
  We estimate the associated electron Lorentz
  factor $\gamma_{\rm cool} \simeq (m_{\rm e} c^2 / \sigma_{\rm T})
  (\Gamma_j / (u_{B} r ))$, following e.g., \cite{Moderski03};
  the corresponding energy rollover in the SED is near
  $\sim 2\times10^{-8} \delta B{\rm [G]} \gamma_{\rm cool}^2 /
  (1+z)$~eV.
Here, $\delta = [\Gamma_{\rm j}(1-\beta_{\rm j}\cos\theta_{\rm obs})]^{-1}$, the
Doppler factor; $\theta_{\rm obs} = \theta_{\rm j}$, the jet
viewing angle; $\beta_{\rm j} = (1-1/\Gamma^2)^{1/2}$, the jet bulk
speed normalized to $c$; and
$u_{\rm B} = B^2/8\pi$ is the magnetic energy density.
  In the case of
  $\Gamma_{\rm j} = 12.5$, a cooling rollover forms in the EUV band
  ($\sim 60\;{\rm eV}$), resulting in a slight deficiency in the soft
  X-ray flux.
For $\Gamma_{\rm j} = 25$, values of $\gamma_{\rm cool}$ span $\sim 2 \times 10^4
- 5 \times 10^4$, with a cooling rollover near $\sim 0.3 - 0.8$~keV, but
this rollover is very mild; spectral curvature in the X-ray band is
dominated by a break in the electron energy distribution.
Our models do not attempt to reproduce the sharp break in the hard X-ray band
suggested by the best-fitting model to the \textit{AstroSAT} data.
We have adopted a maximum electron
energy fixed at $\gamma_{\rm max} = 3\times 10^6$ in order to produce
enough VHE emission.  This results in the synchrotron components
extending almost to the MeV band.
Results for models with
 $\Gamma_{\rm j} = 25$ are listed in Table~\ref{tab:tab_sed_params}.
  %
In Table \ref{tab:tab_sed_params}, we report the expected variability
time-scale $t_{\rm var} = (1+z)R_{\rm j}/{\delta}c$.
The values of $t_{\rm var}$ range from $0.11\;{\rm hr}$ in the $\Gamma_{\rm j} = 25$ case to $1.3\;{\rm hr}$ in the $\Gamma_{\rm j} = 12.5$ case.

We also report three components of the predicted
jet power contained in the magnetic fields, electrons and protons,
assuming one cold proton for each electron \citep[alternatively, assuming
the presence of electron-positron pairs would reduce the predicted
proton jet power, e.g.,][]{Madejski16}.  We find
very consistent results with $L_{\rm B} \ll L_{\rm e} \ll L_{\rm
  p}$. The reason for this consistency is that the mean electron
energy is determined mainly by the low-energy part of their
distribution, which has been assumed to be fixed.  The low energetic
contribution of magnetic fields suggests a matter-dominated jet, which
is a typical result for BL~Lac type objects \citep{Tavecchio16}.

\begin{table}
\caption{Parameters of three SED models:
$\Gamma_{\rm j}$ is the jet bulk Lorentz factor;
$\theta_{\rm j} = 0.2/\Gamma_{\rm j}$ is the jet half-opening angle;
$\theta_{\rm obs} = \theta_{\rm j}$, the jet viewing angle;
$\delta = [\Gamma_{\rm j}(1-\beta_{\rm j}\cos\theta_{\rm obs})]^{-1}$ is the Doppler factor, with $\beta_{\rm j} = (1-1/\Gamma^2)^{1/2}$, the jet bulk speed normalized to $c$;
$r$ is the distance scale of the emitting region along a conical jet;
  $t_{\rm var} = (1+z)R_{\rm j}/{\delta}c$, the expected variability time-scale, with $R_{\rm j} = r\theta_{\rm j}$, the jet radius;
and $p_1, p_2, \gamma_{\rm br}, \gamma_{\rm max}$ are parameters of the
  broken power-law electron energy distribution $N(\gamma) \propto
  \gamma^{-p_1}$ for $1 < \gamma < \gamma_{\rm br}$ and $N(\gamma)
  \propto \gamma^{-p_2}$ for $\gamma_{\rm br} < \gamma < \gamma_{\rm
    max}$.
  Also reported are estimates of the jet power contained in the magnetic fields ($L_{\rm B}$), the electrons ($L_{\rm e}$), and the protons ($L_{\rm p}$, one cold proton per electron).}
\label{tab:tab_sed_params}
\vskip 1ex
\centering
\begin{tabular}{ccccc}
\hline\hline
$\Gamma_{\rm j}$ & 25 & 25 & 25 \\
$B\;{\rm [G]}$ & 0.3 & 0.6 & 1.2 \\
\hline
$\theta_{\rm j} = \theta_{\rm obs}$ & 0.008 & 0.008 & 0.008 \\
$\delta$ & 48 & 48 & 48 \\
$r\;{\rm [pc]}$ & 0.052 & 0.023 & 0.010 \\
$t_{\rm var}\;{\rm [h]}$ & 0.25 & 0.11 & 0.05 \\
$p_1$ & 2.0 & 2.0 & 2.0 \\
$p_2$ & 2.85 & 2.8 & 2.8 \\
$\gamma_{\rm br}\;[\times 10^4]$ & 4.6 & 3.6 & 3.2 \\
$\gamma_{\rm max}\;[\times 10^6]$ & 3.0 & 3.0 & 3.0 \\
$\log L_{\rm B}\;{\rm [erg\,s^{-1}]}$ & 41.5 & 41.5 & 41.3 \\
$\log L_{\rm e}\;{\rm [erg\,s^{-1}]}$ & 43.1 & 43.0 & 42.8 \\
$\log L_{\rm p}\;{\rm [erg\,s^{-1}]}$ & 45.3 & 45.2 & 45.1 \\
\hline\hline
\end{tabular}
\end{table}

\section{Discussion}


\subsection{Basic constraints from continuum variability}


We conducted a 4-day contemporaneous campaign in X-ray, TeV, and
optical bands on the HBL Mkn 421 to explore its short-term variability
properties.  One of our campaign's strengths is the energy-resolved
X-ray light curves from {\it AstroSAT} SXT, allowing us to examine
X-ray variability on time-scales down to a few ks.  The campaign caught
four small flares in the X-ray band, each lasting roughly 50--100\,ks.

An observed variability timescale can be used to infer an upper limit
estimate for the size $R$ of the emission region in the co-moving frame
based on light travel time, and ignoring contributions from dynamic
processes, via $ R < c t_{\rm var} \delta (1+z)^{-1} $.  We first
consider $t_{\rm var}$ to be the fastest observed variability
time-scale.  During our observation, the source X-ray flux increased
by 22\,per cent in six orbits (35.0\,ks) in the 0.6--0.8\,keV band
(starting roughly 60 ks after the start of the \textit{AstroSat}
observation), and it increased by 82\,per cent in two orbits
(11.7\,ks) in the 3--7\,keV band (starting at roughly 70\,ks after the
start).  Adopting $\delta=25$ hereafter, a value consistent with the
range derived from various SED fits \citep[e.g., ][]{Maraschi99,
  Abdo11,Aleksic15a,Aleksic15b,Acciari20} or rapid TeV variability
\citep{Celotti98}, these values of $t_{\rm var}$ yield $R < 3 \times
10^{16}$ cm and $R < 8 \times 10^{15}$ cm for the 0.6--0.8\,keV and
3--7\,keV bands, respectively.
Using a value of black hole mass $M_{\rm BH} = 1.9\times10^8 \Msun$
\citep{WooUrry02}, 1~$R_{\rm g}$ = $ 2.9 \times 10^{13}$\,cm, and thus
$R < 1000 R_{\rm g}$ or $R < 300 R_{\rm g}$ for the 0.6--0.8 or
3--7\,keV bands, respectively.
Alternatively, one can use the full duration of the flares as the
variability time-scale, neglecting blending between successive flares.
Using the same start/stop times as defined in $\S$\ref{sec:fitflare}
(Table~\ref{tab:FLARETABLE}), the average start-to-stop flare duration
is 73 ks in the 0.6--0.8\,keV band and 69 ks in the 3--7\,keV band.
Using these as $t_{\rm var}$ yields $R < 5 \times 10^{16} {\rm cm} \sim 1800~R_{\rm g}$.

A scenario in which only cooling dominates the variability is
excluded: we would expect relatively higher-energy X-ray bands to fall
more quickly, but as is apparent from Fig.~\ref{fig:SXTlcoverplot},
all bands fall at roughly similar rates.

We can estimate the co-moving magnetic field strength $B$ based on observed X-ray flux decays,
and constraining the synchrotron cooling time-scale to be shorter
than the fastest variability time-scale.
The fastest observed decay time in the X-rays is in the 3--7\,keV band, over 17.5\,ks (3 orbits).
We start from the synchrotron power emitted \citep{RL79}, 
$P_{\rm syn} = \frac{4}{3} \sigma_{\rm T} c \beta^2 \gamma^2 u_{\rm B}$, where
$u_{\rm B} = B^2/8\pi$ is the magnetic energy density,
$\sigma_{\rm T}$ is the Thomson cross section, $6.7\times10^{-25}$~cm$^{-2}$,
$\gamma$ is the Lorentz factor for the emitting electron,
$\beta \equiv v/c \sim 1$ for ultra-relativistic electrons.
We define the synchrotron cooling time as $\tau_{\rm synch} \equiv$
(electron energy)/(synchrotron power loss) = $\gamma m_{\rm e} c^2 /
P_{\rm syn}$, or
\begin{equation}
\tau_{\rm synch} = \frac{3 m_{\rm e} c}{4 \sigma_{\rm T} \gamma u_{\rm B}}
\end{equation}
Note this refers to the cooling time in the rest frame; the observed
cooling time-scale $\tau_{\rm synch,obs} = \tau_{\rm synch} (1+z) \delta^{-1}$.
Solving for $B$, we obtain
\begin{equation}
  B = \sqrt{\frac{6\pi ~ (1+z) ~ m_{\rm e} c}  {\sigma_{\rm T} ~ \delta ~ \gamma ~ \tau_{\rm synch,obs}}}
\end{equation}
To solve for $\gamma$, we use the expression
for the peak synchrotron emission energy
produced by monoenergetic electrons with Lorentz factor $\gamma$ and averaged over
magnetic pitch angle from \citet{Nalewajko11},
\begin{equation}
  E_{\rm syn}  =  2\times10^{-8} \delta B \gamma^2 (1+z)^{-1} {\rm eV }
  \label{eq:KN11Egam}
\end{equation}
Rearranging to solve for $B$, we obtain
\begin{equation}
B =  \left(\frac{6\pi ~ m_{\rm e} ~ c}{\sigma_{\rm T} ~ \tau_{\rm synch,obs}}\right)^{2/3}\left(\frac{(2\times10^{-8}{\rm eV}) (1+z)}{\delta ~ E_{\rm syn}}\right)^{1/3} {\rm G}
  \end{equation}
Using $E_{\rm syn}$=5000 eV, and using the average 3--7~keV flare duration
as an upper limit to $\tau_{\rm synch,obs}$,
we obtain a lower limit on the co-moving magnetic field: $B > \sim 0.08 \delta^{-1/3}$ G.  
Defining $\delta_{25} \equiv \delta/25$, therefore
$\delta^{-1/3} = 0.3\times\delta_{25}^{-1/3}$, and finally
$B > \sim 0.02 \delta_{25}^{-1/3}$ G.



\subsection{Constraints from intra-X-ray cross-correlations}

During the first two X-ray flares observed in our campaign, relatively
harder bands increase in flux later than softer bands; it can be seen
from Fig.~1 that during
the first two minima the 0.6--0.8\,keV band begins to increase
while the the 3--7\,keV band is still decreasing. These trends drive a
soft-to-hard lag between the 0.6--0.8\,keV band and the other bands.
However, the harder bands increase more rapidly, and even peak earlier
than the softer bands.  These observations are consistent with the
notion that when there is a dissipation event, it develops at
relatively lower energies first, but works more effectively towards
harder bands later on. In contrast, the third and fourth flares are
more time-symmetric across all X-ray sub-bands.

The fact that most bands are well-correlated at lags consistent with
zero supports emission originating in homogeneous regions where particles emitting
high- and low-energy photons are co-spatial. 
It could also indicate that cooling and acceleration time-scales are
overall roughly equal, particularly for flares 3 and 4.  (Having
$\tau_{\rm cool}$ $\sim$ $\tau_{\rm accel}$ is reasonable to expect
given that the X-ray band in HBLs corresponds to the high-energy
regime of synchrotron emission; \citealt{Kirk98}). 

However, cross-correlations do not cleanly distinguish between flares
peaking first at harder energies versus acceleration impacting the
softer bands first.  Both processes seem to be present during flares 1
and 2; the combination of processes can lead to overall observed lags
near zero.

Nonetheless, given that we do measure a net soft-to-hard lag across
flares 1 and 2, we can explore further the notion that acceleration
processes impact relatively lower energies first, so that
energized electrons appear at
low energies first and gradually build up towards higher energies
\citep{GM98}.

Under the assumption that cooling and acceleration time-scales are
roughly equal, the soft-to-hard lag indicates acceleration occurring
at lower electron energies first and progressively later to higher
energies following Eqn.~9 of \citet{Zhang02}:

\begin{equation}
B~\delta~\xi_{\rm acc}^{-2/3} = 0.21 ~ (1+z) ~ E_{\rm hi}^{1/3} ~ \left[\frac{1 - (E_{\rm lo}/E_{\rm hi})^{1/2}}{\tau_{\rm lag}}\right]^{2/3}  ~ {\rm G}
\end{equation}


where $B$ is the co-moving magnetic field strength in Gauss, and $\tau_{\rm lag}$ is the value of the
observed soft-to-hard delay from $E_{\rm lo}$ to $E_{\rm hi}$ in
seconds (best-fitting value of 4600).  Using the pitch angle-averaged
photon energy as in Eq.~\ref{eq:KN11Egam} introduces a mere 10\,per cent
difference (one replaces the constant 0.21 with 0.19).  We adopt 0.7
and 2\,keV for $E_{\rm lo}$ and $E_{\rm hi}$, respectively, and again adopt
$\delta_{25} \equiv \delta/25$.

As discussed in $\S$5.2 of \citet{Zhang02}, the
value of the acceleration parameter $\xi_{\rm acc}$ is not clear; we
define $\xi_{\rm acc,5} \equiv \xi_{\rm acc} / 10^5$.  We obtain
$B~\sim~0.05 ~ \delta_{25}^{-1} ~ \xi_{\rm acc,5}^{2/3}$ ~ G.  However, given that the
observed lag has likely been diluted by contemporaneous hard-to-soft
processes, this derived estimate of $B$ can be treated as a lower
limit.

Hard-to-soft lags are generally expected in the context of the
electron population becoming softer due to synchrotron radiative
losses; if synchrotron losses dominate,
then the magnitude of the hard-to-soft lag can constrain $\tau_{\rm
  cool}$, and thus $B$, following e.g., \citet[][their eqn.~1]{Chiappetti99}.

In this campaign, we find only upper limits on hard-to-soft lags; the
highest upper limit was 4.4\,ks (ICF centroid, second half only, 3--7\,keV
band leading 0.6--0.8\,keV), which implies a lower limit of
$B{\delta}^{1/3} \gtrsim 0.7$\,G following \citet{Chiappetti99}, or
$B {\delta_{25}^{1/3}} \gtrsim 0.2$\,G.


\subsection{Constraints from energy-dependent variability amplitudes}

We observed fractional variability amplitude $F_{\rm var}$ to increase
with energy across the synchrotron hump, following a best-fitting relation
$F_{\rm var} \propto E^{+0.20}$ across both the X-ray band and from
optical/NIR to the X-rays, and consistent with previous results for
Mkn~421.

An observed energy dependence for $F_{\rm var}$ argues against
variability being due predominantly to changes in magnetic field
strength or size of emission region or due to changes in Doppler
factor $\delta$ (e.g., from viewing angle). That is, the
    observed flares' being produced by rapid precession of a single
    emission component about the jet axis is unlikely, despite the
    roughly similar amplitudes of all four flares.   The observed
energy dependence instead supports variability being dominated by
acceleration/cooling.

An alternate possibility is that the occurrence of relatively lower
fractional variability towards lower energies could be consistent with
an additional constant or less-variable emission term: assuming a
variability process that yields the same variance across a waveband,
increasing the mean by a certain factor would reduce $F_{\rm var}$ by
that factor.  An emission component whose integrated 0.6--0.8\,keV flux
is twice that in the 3--7\,keV band would have values of $F_{\rm var}$
in those bands differing by 2, as observed.  Conjecturally
extrapolating to the optical band, one requires an emission component
with flux integrated across a given optical filter $\sim8$ times that
at 3--7\,keV. However, for the sake of exploring models in a
straightforward manner, we ignore such a putative term and assume that
all observed emission is from variable components of the jet.

We can consider a context in which the observed synchrotron emission
is the superposition of radiation from a large number of distinct
individually-emitting cells.  We follow the turbulent cells model of
\citet{Marscher10}, in which cells have a similar size,
and their EED widths are distributed such that those cells
capable of emitting at relatively higher energies are present in
relatively smaller volume filling factors.
More specifically, we consider a phenomenological toy model in which
each cell emits a spectrum represented by a power-law with photon
index $\Gamma_0=1.3$, modified with a high-energy exponential rollover
at energy $E_{\rm cut}$.  As a starting assumption,
we assume that this power-law extrapolates from X-rays (10\,keV)
down to the optical/IR band (1 eV). The number density of cells that
have a certain value of $E_{\rm cut}$ follows a power-law distribution
as d$N$/d$E_{\rm cut}$ $\propto$ $E_{\rm cut}^{-\alpha_0}$.
We consider a model set-up with $N$ such cells; typically of order
30000 cells are needed for numerical convergence and to yield
variability amplitudes consistent with our observations. We assume that
all cells have the same size, and are all subject to identical amounts
of radiative cooling and Doppler beaming.

\begin{figure}
  \centering
  \includegraphics[width=0.99\columnwidth]{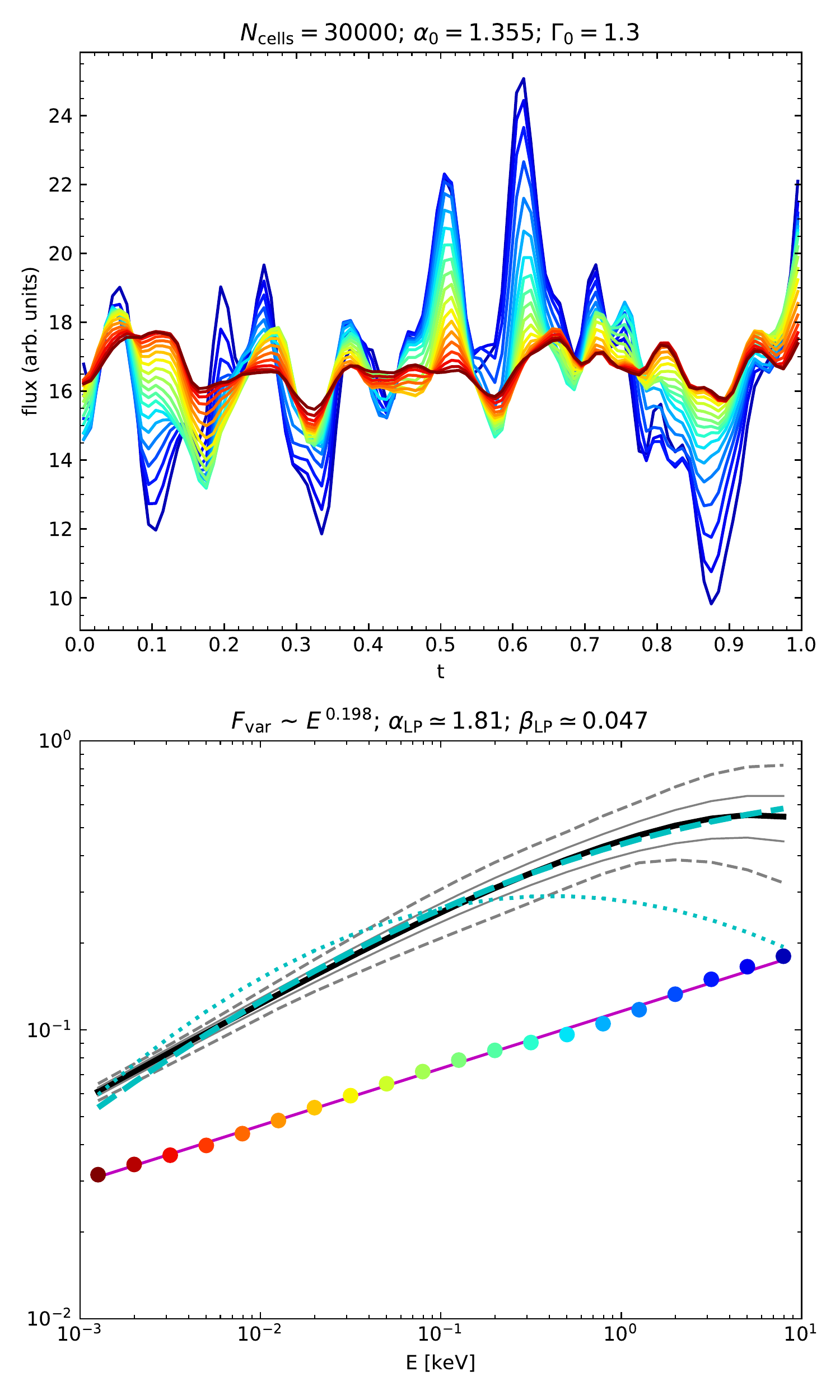}
  \caption{Results from one run of our toy cell model
based on having a large number of independent emission sources, with emission
characterized by a power-law distribution of exponential cut-off energies
following d$N$/d$E_{\rm cut}$ $\propto$ $E_{\rm cut}^{-\alpha_0}$.
The top panel shows a simulated set of 20 energy-dependent light curves
from optical (dark red; down to 1 eV) to X-ray (dark blue; up to 10\,keV)
energies, all normalized to the same mean. 
In this particular run, $\alpha_0=1.355$.
The bottom panel shows the variability amplitudes $F_{\rm var}$,
with colours corresponding to the light curves displayed in the left panel.
The solid magenta line shows the best-fitting power-law trend of
$F_{\rm var} \propto E^{0.198}$. The solid black line shows the mean (averaged over
time) $\nu F_\nu$ SED (in arbitrary units) obtained in the simulation,
the solid gray lines indicate the $\pm 1$$\sigma$ confidence limits, and
the dashed gray lines indicate the mininum and maximum values. The dashed cyan
line is the best log-parabola model (with parameters $\alpha_{\rm LP}
\simeq 1.81$ and $\beta_{\rm LP} \simeq 0.047$) fitted to the solid
black line. The dotted cyan line is the log-parabola model that
matches the observed optical-to-X-ray SED of Mkn~421 (with arbitrary
normalization).
  }
    \label{fig:KNcells141}
  \end{figure}

\begin{figure} \centering
  \includegraphics[width=0.9\columnwidth]{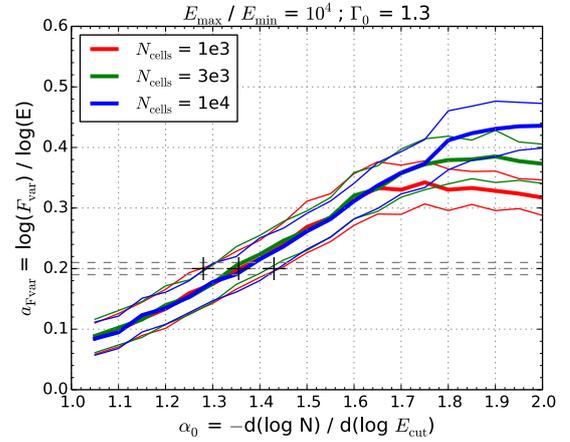}
  \caption{Results of simulations detailing the behaviour of $a_{\rm
      Fvar}$, the power-law index of variability amplitude as function
    of observed photon energy, as a function of $\alpha_0$, which
    quantifies the distribution of high-energy spectral cutoffs.  Red,
    green, and blue lines indicate simulations with 3000, 10000, and
    30000 cells, respectively.  For each value of $\alpha_0$, the
    thick and thin solid lines indicate the mean and standard
    deviations, respectively, of 100 simulations. Values of $\alpha_0$
    between 1.28 and 1.43 yield values of $a_{\rm Fvar}$ consistent
    with the best-fitting observed value of $a_{\rm Fvar}$.   }
    \label{fig:KNcells142}
  \end{figure}

We simulate light curves with 100 time steps, and consisting of
multiple time-symmetric flares: exponentially-rising and -decaying
flares are generated following Eqn.~\ref{eq:flareshape}, with the time
between successive flares chosen from uniform
distributions. For simplicity,
rise and fall times are set equal to each other at all energies
and to 1/100 of the light curve duration.
We simulate light curves in 20 energy sub-bands logarithmically spaced
from 1~eV to 10~keV.  For simplicity, we neglect the energy-dependency
of synchrotron cooling times.  An example set of energy-dependent
light curves is plotted in Fig.~\ref{fig:KNcells141}.  We sum the
emission in each sub-band, measure $F_{\rm var}$ from each of the 20
summed sub-band light curves, and fit a power-law model to $F_{\rm
  var}$($E$) $\propto E^{-a_{\rm Fv}}$ as plotted in the second panel
of Fig.~\ref{fig:KNcells141}.  We tested values of $\alpha_0$ ranging
from 1.05 to above 2.0 in steps of 0.05, though solutions tended to
not converge above $\alpha_0=2.0$.  We repeated this procedure 100
times to build up distributions of $a_{\rm Fv}$ for each value of
$\alpha_0$ tested. As plotted in Fig.~\ref{fig:KNcells142}, we find
that $\alpha_0=1.35^{+0.08}_{-0.07}$ yields results consistent with
the observed value of $a_{\rm Fv} = 0.20\pm0.01$.


This is admittedly a simple toy model, and there is room for
additional development; one could additionally consider variations in
source size, Doppler beaming, radiative cooling, and in d$N$/d$E_{\rm
  cut}$. Such considerations could potentially lead to the model
output broadband energy spectrum having more curvature at 0.1--1\,keV,
thus more closely matching the broadband shape of the observed
SED. However, we leave such detailed development for future papers.

\section{Conclusions}

We have presented results obtained from a 4-day coordinated
multi-wavelength campaign on the HBL Mkn~421 in January 2019,
conducted to better study this blazar's short time-scale variability
properties and gain insight into the properties of the emitting regions.

We gathered continuous X-ray monitoring with \textit{AstroSat} SXT and
LAXPC, tracking full- and sub-band variability down to satellite
orbital time-scales (98 minutes).  The FACT telescope provided nightly
TeV monitoring, with up to $\sim22$\,ks ($\sim6$\,hours) of data taken each night.
The Whole Earth Blazar Telescope consortium (most notably the EPT
telescope on La Palma) provided contemporaneous B-, V-, R-, and I-band
photometric monitoring during our campaign.

Our primary findings are summarized as follows:

\begin{itemize}

\item Our campaign caught four modest X-ray flares ($\sim20-60$ per cent flux
  increases/decreases, depending on sub-band energy), each lasting on
  time-scales of roughly $\lesssim$ a day. However, the source
  exhibited minimal night-to-night variability in any energy range, precluding
  any meaningful interband correlation analysis from this campaign.

\item We studied energy-dependent variability within the X-ray band
  using \textit{AstroSat} SXT. During the first two flares, relatively
  harder X-rays are associated with shorter rise times and longer
  decay times, while relatively softer X-rays are associated with more
  symmetric flares.  The relatively harder bands increase in flux
  later and faster compared to the softer X-rays, leading to an
  observed soft-to-hard lag of an average of $4.6 \pm 2.6$\,ks from the
  0.6--0.8\,keV sub-band to the harder sub-bands during the first two
  flares (though there likely co-exist contemporaneous hard-to-soft
  and soft-to-hard emission processes). The third and fourth flares
  are more time-symmetric across all X-ray sub-bands.

\item Continuum flux variability amplitudes increase monotonically with energy across
  both the X-ray band ($F_{\rm var}$ $\propto$ $E^{0.26\pm0.01}$) and
  across the synchrotron hump from NIR/optical to X-rays ($F_{\rm
    var}$ $\propto$ $E^{0.20\pm0.01}$), consistent with previous
  measurements for Mkn~421 and for other HBL blazars.  We explain this
  behaviour in the context of a simple toy model in which the observed
  synchrotron emission originates in a large number of individual
  cells whose EEDs are distributed such that relatively higher
  energy-emitting cells occupy relatively smaller volumes.  Each cell
  emits a power-law spectrum across the NIR/optical-to-X-ray band, but
  with an exponential cutoff, and the distribution of cutoffs $E_{\rm
    cut}$ follows a power law as d$N$/d$E_{\rm cut}$ $\propto$ $E_{\rm
    cut}^{-\alpha_0}$.  Simulations of sub-band light curves that
  follow such a distribution yields the result that a power-law index
  of $\alpha_0=1.35^{+0.08}_{-0.07}$ yields results an
  energy-dependent variability behaviour consistent that observed
  during our campaign.

 \item We fit the time-averaged X-ray spectrum, composed of 0.6--7\,keV
   SXT and 3--7\,keV LAXPC data.  A mildly broken power law (with
   photon index $\Gamma_1 = 2.20 \pm 0.01$ below a break energy of
   $5.0^{+0.2}_{-0.2}$~keV breaking to $\Gamma_2 =
   2.73^{+0.23}_{-0.13}$ above it) is a superior fit than an
   unbroken power-law.  Time-resolved fits to orbitally-binned data
   using this broken power-law model, and with $\Gamma_2$ fixed to
   $\Gamma_1 + 0.49$ yields the usual qualitative
   hardening-when-brightening behaviour common to blazar emission.  No
   obvious signs of hysteresis in the $\Gamma_1$--flux plane were
   observed during our campaign; no signs of differing spectral
   variability behaviour between the four X-ray flares was evident.

  \item We constructed the broadband SED for this campaign, combining
    \text{AstroSat}, optical/NIR flux densities, FACT TeV flux
    densities, and contemporaneous \textit{Fermi} LAT data.  A
    standard one-zone leptonic model including synchrotron and SSC
    emission fits the overall SED well for a co-moving magnetic field strength $B \sim 0.3 \delta_{25}$~G.
    The model fits imply a low magnetic power and
    a matter-dominated jet, as is typically found for BL~Lac type objects.

  \item
    We have derived multiple constraints on the co-moving magnetic field strength:
    $B > \sim 0.02 \delta_{25}^{-1/3}$~ G by requiring the synchrotron cooling time-scale 
    to be shorter than the typical flare time-scale ($\S$8.1);
    $B > \sim 0.05 \delta_{25}^{-1} \xi_{\rm acc,5}^{2/3}$~ G from observed soft-to-hard intra-X-ray lags ($\S$8.2); and
    $B > 0.2 \delta_{25}^{-1/3}$~ G based on upper limits to hard-to-soft lags ($\S$8.2). These limits are consistent with
    the estimate of  $B   \sim 0.3 \delta_{25}$~ G obtained from  SED model fits ($\S$7).

    
  \end{itemize}


\section*{Acknowledgements}


A.G.M.\ and S.K.\ acknowledge partial support from Narodowe Centrum 
Nauki (NCN) grants 2016/23/B/ST9/03123 and 2018/31/G/ST9/03224.
A.G.M.\ also acknowledges support from NCN grant 2019/35/B/ST9/03944.
K.N.\ was supported by NCN grant 2015/18/E/ST9/00580.
G.B.\ acknowledges support from NCN grant 2017/26/D/ST9/01178.
S.Z.\ acknowledges support from NCN grant 2018/29/B/ST9/01793.
E.B.\ acknowledges support from DGAPA-UNAM grant IN113320.
S.M.\ Hu acknowledges the support from the Natural Science Foundation of China under grant No.\ 11873035.
K.M.\ acknowledges support from JSPS KAKENHI grant number 19K03930.

This publication uses the data from the AstroSat mission of the Indian
Space Research Organisation (ISRO), archived at the Indian Space
Science Data Centre (ISSDC).
This work has used the data from the Soft X-ray Telescope (SXT)
developed at TIFR, Mumbai, and the SXT POC at TIFR is thanked for
verifying and releasing the data via the ISSDC data archive and
providing the necessary software tools.

This research has made use of ISIS functions (ISISscripts) provided by
ECAP/Remeis observatory and MIT
(http://www.sternwarte.uni-erlangen.de/isis/).

This research has made use of data and/or software provided by the
High Energy Astrophysics Science Archive Research Center (HEASARC),
which is a service of the Astrophysics Science Division at NASA/GSFC.

The important contributions from ETH Zurich grants ETH-10.08-2 and
ETH-27.12-1 as well as the funding by the Swiss SNF and the German
BMBF (Verbundforschung Astro- und Astroteilchenphysik) and HAP
(Helmoltz Alliance for Astroparticle Physics) are gratefully
acknowledged. Part of this work is supported by Deutsche
Forschungsgemeinschaft (DFG) within the Collaborative Research Center
SFB 876 "Providing Information by Resource-Constrained Analysis",
project C3. We are thankful for the very valuable contributions from
E.\ Lorenz, D.\ Renker and G.\ Viertel during the early phase of the
project. We thank the Instituto de Astrof\'isica de Canarias for
allowing us to operate the telescope at the Observatorio del Roque de
los Muchachos in La Palma, the Max-Planck-Institut f\"ur Physik for
providing us with the mount of the former HEGRA CT3 telescope, and the
MAGIC collaboration for their support.

This research was partially supported by the Bulgarian National
Science Fund of the Ministry of Education and Science under grants DN
18-10/2017, DN 18-13/2017, KP-06-H28/3 (2018), KP-06-H38/4 (2019) and
KP-06-KITAJ/2 (2020). 

Part of the photometric data included in this work
were collected during the photometric monitoring observations with the
robotic and remotely controlled observatory at the University of
Athens Observatory \citep[UOAO;][]{Gazeas2016}.

The authors (Dr.\ J.R.\ Webb) are honored to be permitted to conduct
astronomical research on Iolkam Du’ag (Kitt Peak), a mountain with
particular significance to the Tohono O’odham Nation.

This work is partially based on observations collected at the
Observatorio Astron\'omico Nacional at San Pedro M\'artir Baja
California, Mexico.

\section*{Data availability statement}

The data underlying this article will be shared on reasonable request
to the corresponding author.






\appendix

\section{\textit{AstroSAT} SXT periodogram measurement}

We measured the periodogram of the 1.4--1.9\,keV SXT light curve; it
had the highest count rate and thus lowest expected power due to
Poisson noise.
Following the method of \citet{Vaughan05}, we first model the broadband
continuum noise by fitting a power-law in log-log space to the
red noise-dominated portion of the periodogram, excluding the candidate bump and
those points above $\sim6\times10^{-5}$\,Hz dominated by Poisson noise.

Our best-fitting power-law slope is $-1.96 \pm 0.14$, and the best-fitting normalization
(logarithm of power in units of rms$^2$~Hz$^{-1}$ at $10^{-5}$~Hz) is $2.16 \pm 0.21$.
The candidate bump at $\sim 1.5\times10^{-5}$\,Hz yields a value of the
ratio $I/P$ (where $I$ denotes the periodogram power of the candidate
signal and $P$ denotes the value of the best-fitting continuum model) of
6.02 for our best-fitting model, but as low as only 3.74 considering the parameter
uncertainties (the relative scarcity of sampled frequencies here means
that the measured continuum level is not well constrained).
Applying Eqn.~16 of \citet{Vaughan05} to take into account the ``look
elsewhere'' effect \citep{Algeri16} given the number of independent frequencies
sampled yields that the bump is inconsistent with being due to pure
red noise at the 93.1\,per cent confidence level for the best-fitting
model, but merely the 48\,per cent confidence level considering the
parameter uncertainties.

In addition, caution must be applied here on multiple levels,
particularly given that pure red noise processes can frequently mimic
few-cycle sinusoid-like quasi-periods, and that we only measure four
``cycles'' here.  We follow the recommendation of \citet{Vaughan16}
that roughly five cycles are typically needed to distinguish between
red noise and a true period.  Finally, fully considering
the ``look elsewhere'' effect means one
should also take into account those X-ray power-spectra of Mkn 421
previously published and where a narrow power component was not
claimed \citep[e.g.,][]{Chatterjee18}.  In conclusion, the observed
quasi-periodic behaviour does not indicate significant deviation from a
pure red-noise process.



\bsp	
\label{lastpage}
\end{document}